\documentclass[11pt,oneolumn,floatfix,altaffilletter,superscriptaddress,tightenlines,showpacs,showkeys,preprintnumbers,nofootinbib]{revtex4-2}
\pdfoutput=1
\usepackage[colorlinks=true,citecolor=blue,linkcolor=blue,breaklinks=true,linktoc=none]{hyperref}
\usepackage{amsmath,amssymb}
\usepackage{epsfig} 
\usepackage{graphicx}
\usepackage{url}
\usepackage{color}
\usepackage{multirow}
\usepackage{placeins}
\usepackage{float}
\usepackage[dvipsnames]{xcolor}
 \usepackage{booktabs} 
\usepackage{braket}
\usepackage{float}
\usepackage{slashed,tabularray}
\hypersetup{
  colorlinks=true,
  linkcolor=red,
  filecolor=magenta,   
  urlcolor=blue,
  citecolor=blue,
}

\hypersetup{colorlinks,linkcolor={blue},citecolor={teal},urlcolor={violet}}

\allowdisplaybreaks

\bibpunct{[}{]}{,}{n}{}{,}

\def\beq{\begin{equation}}
\def\eeq{\end{equation}}
\def\bea{\begin{eqnarray}}
\def\eea{\end{eqnarray}}
\makeatletter
\@addtoreset{equation}{section}

\newcommand{\INFN}{INFN - Sezione di Napoli, Complesso Universitario Monte S. Angelo, I-80126 Napoli, Italy}

\newcommand{\SSM}{Scuola Superiore Meridionale, Università degli studi di Napoli ``Federico II'', Largo San Marcellino 10, 80138 Napoli, Italy}
\newcommand{\NAN}{Department of Physics and Institute of Theoretical Physics,
Nanjing Normal University, Nanjing, 210023, China}
\newcommand{\JSI}{Jo\v{z}ef Stefan Institute, Jamova 39, 1000 Ljubljana, Slovenia}

\begin{document}

\title{Mapping Domain-Wall Bayesian Reconstruction with LISA} 
\author{Satyabrata Datta}
    \email{amisatyabrata703@gmail.com}
    \affiliation{\JSI}
    \affiliation{\NAN}
 \author{Rome Samanta}
  \email{samanta@na.infn.it}
  \affiliation{\SSM}
  \affiliation{\INFN}

\begin{abstract}
We study the Bayesian reconstruction of peaked domain-wall
gravitational-wave signals at LISA and construct reconstruction
maps over the signal-parameter plane. These maps identify
the regions in which the signal can be probed with minimal
posterior uncertainty and parameter degeneracy.
Our analysis employs a two-parameter domain-wall spectral
template and includes isotropic, unmodulated astrophysical
foregrounds from Galactic double-white-dwarf binaries and
extragalactic compact binaries, together with LISA
instrumental noise. The inference is performed for 64
injection points distributed on an equidistant grid using
nested sampling, and the resulting posterior quantities
are interpolated with the Clough--Tocher method to generate
smooth maps over the full parameter plane.
Under the adopted instrumental-noise and astrophysical-foreground
modelling assumptions, we find that LISA reconstructs
domain-wall signals most effectively when the annihilation
temperature lies approximately in the range
$10^3\text{--}10^6\,\mathrm{GeV}$.
In this regime, the posterior becomes both tighter and
less degenerate, enabling genuine two-parameter reconstruction.
The most favorable region corresponds to signals with
${\rm SNR}\gtrsim 50$, while signals with ${\rm SNR}\sim 10$
can still be reconstructed effectively only in a narrower
part of parameter space concentrated near
$T_*\sim 10^{5-6}\,\mathrm{GeV}$.
In terms of the observable spectrum, this weaker-signal region corresponds approximately to peak amplitudes in the range $4\times10^{-12}\lesssim \Omega_{\rm GW}^{\rm peak}h^2 \lesssim 4\times10^{-11}$, with characteristic peak frequencies satisfying $30\,{\rm mHz}\lesssim f_p\lesssim 100\,{\rm mHz}$.
Our results provide a quantitative reconstruction forecast
for peaked domain-wall signals in the LISA band and a useful
guide for particle-physics realizations of domain walls
that predict peaked gravitational-wave spectra in the
milli-Hz range.
\end{abstract}
\maketitle
\tableofcontents
\section{Introduction}

The direct detection of gravitational waves (GWs) by LIGO opened a new observational window on the Universe and established GW astronomy as a precision probe of fundamental physics and astrophysics \cite{LIGOScientific:2016aoc}. Since then, the field has expanded rapidly across a broad frequency range. Ground-based interferometers probe the audio band, pulsar timing arrays access nanohertz frequencies, and space-based observatories such as LISA are designed to explore the milli-Hz band \cite{LIGOScientific:2016jlg,lisa,ng1,ng2,ng3,ng4,ng5}. In parallel, future facilities promise broader frequency coverage and improved sensitivity \cite{et,decigo}. In this sense, GW astronomy is entering an era in which cosmological source classes can be studied not only through detectability forecasts but also through detailed parameter-reconstruction analyses.

Among proposed cosmological GW sources, topological defects are especially interesting because they naturally generate spectra extending over many decades in frequency. In particular, domain walls can source a stochastic GW background whose shape is governed by a small number of physically meaningful parameters related to the wall energy density and annihilation scale \cite{Zeldovich:1974uw,Press:1989yh,Saikawa:2017hiv}. This makes them attractive targets not only for discovery forecasts but also for inference studies: if such a signal is observed, one would like to know under what conditions the underlying domain-wall parameters can actually be reconstructed from the data.

A key feature of the domain-wall spectrum is that its reconstruction properties depend sensitively on the location of the spectral peak relative to the detector band. If the peak lies well outside the LISA window, the detector probes only one asymptotic side of the spectrum, and the data constrain mainly an effective parameter combination. In that regime, the posterior is expected to become strongly elongated in the original parameter basis. By contrast, if the peak lies within or near the most sensitive part of the LISA band, the detector becomes sensitive not only to the overall signal amplitude but also to the spectral turnover and local curvature. One may then expect the degeneracy to be significantly reduced and, in the most favorable part of parameter space, genuine two-parameter reconstruction to become possible.

This is the regime studied in the present work. Our goal is to map the reconstruction landscape of peaked domain-wall signals in the LISA band and to identify where in parameter space the signal can be reconstructed with both small uncertainty and weak degeneracy. This leads to several natural questions: where in the signal-parameter plane is the posterior least degenerate, where are the reconstruction uncertainties smallest, and do these two conditions coincide? Addressing these questions requires going beyond simple detectability estimates.

A realistic assessment of this problem must account not only for instrumental noise, but also for the astrophysical foreground environment in the milli-Hz band. In particular, the unresolved Galactic population of double white-dwarf binaries is expected to generate a confusion foreground that dominates a substantial part of the LISA band and constitutes the leading obstacle to the extraction of weak cosmological backgrounds \cite{Adams:2013qma,Boileau:2020rpg,Boileau:2021sni,Korol:2021pun,Korol:2020lpq,Liu:2023qap}. In addition, extra-galactic compact binaries produce a smoother stochastic background that can further complicate the interpretation of a cosmological signal \cite{Phinney:2001di,Regimbau:2011rp,Babak:2023lro,Lehoucq:2023zlt}. Any meaningful reconstruction forecast for domain-wall signals in the LISA band must therefore be carried out in the presence of both instrumental noise and astrophysical foreground contamination.

A further important point is also methodological. In degenerate Bayesian inference problems, commonly used summaries such as one-dimensional marginal widths or total posterior area can be misleading. A posterior may be broad but nearly isotropic, or narrow but strongly elongated, and these two situations have very different physical interpretations. To disentangle these effects, it is useful to characterize the posterior through its principal axes. In particular, the ratio of the principal variances provides a direct measure of degeneracy, while the variance along the short axis isolates the genuinely constrained parameter combination. This principal-component viewpoint is therefore especially well suited to the present problem, where the transition from tail-dominated inference to peak-dominated inference is expected to be encoded directly in the geometry of the posterior.

Our analysis is based on a 10-dimensional Bayesian inference problem consisting of two domain-wall signal parameters and eight nuisance parameters describing instrumental noise and astrophysical foregrounds. We consider 64 injections distributed on an equidistant grid in the signal-parameter plane and perform the inference with nested sampling. The resulting posterior summaries are then interpolated using the Clough--Tocher method \cite{LeeSchachter1980,RenkaCline1984,Nielson1983} to construct smooth reconstruction maps over the full parameter plane. These maps allow us to identify not only where the signal is detectable, but also where the posterior is simultaneously tight and weakly degenerate.
\begin{figure}
    \centering
    \includegraphics[width=0.7\linewidth]{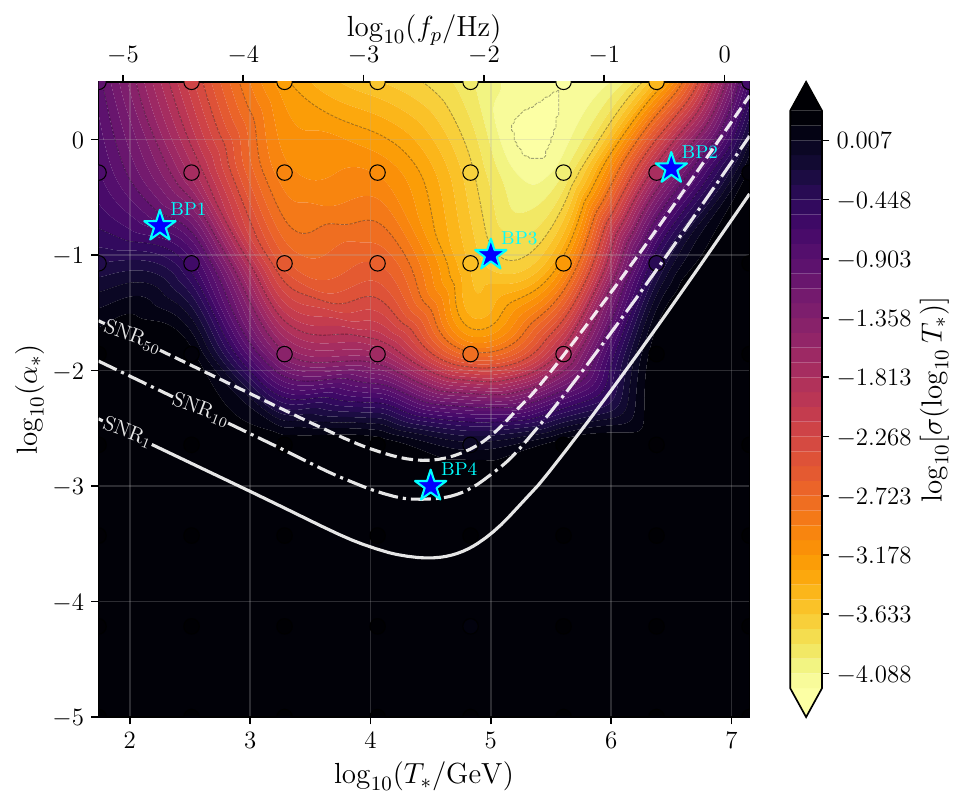}
   \caption{Representative heat map of $\sigma(\log_{10}T_*)$, which measures the reconstruction uncertainty in the annihilation temperature and, equivalently, in the peak frequency, since $f_p\propto T_*$. The map is obtained using the domain-wall signal template defined in Eq.~\eqref{eq:omega_dw}, marginalized over the isotropic Galactic foreground from double white-dwarf binaries and the isotropic extra-galactic foreground from compact-binary coalescences, modeled by Eqs.~\eqref{eq:galac} and \eqref{eq:exgalac}, respectively. In the central region, typically corresponding to $T_*\sim 10^3\text{--}10^6\,\mathrm{GeV}$, values satisfying $\log_{10}[\sigma(\log_{10}T_*)]\lesssim -1$ indicate approximately better than $25\%$ reconstruction of the annihilation temperature, and hence of the peak frequency. Toward the upper edges of the map, however, the posterior becomes strongly degenerate, so the marginalized widths alone cease to provide the most faithful reconstruction metric, as discussed in the main text. The same heat map is shown again in the bottom panel of Fig.~\ref{fig:fig1}, together with the corresponding map of $\sigma(\log_{10}\alpha_*)$, where $\alpha_*$ denotes the fractional energy density stored in domain walls at the annihilation temperature $T_*$. For the latter, the corresponding threshold for approximately $25\%$ reconstruction of the peak amplitude, using $\Omega_{\rm GW}^{\rm peak}\propto \alpha_*^2$, is $\log_{10}[\sigma(\log_{10}\alpha_*)]\lesssim -1.31$. The four-starred benchmark points are used to generate the corner plots in Figs.~\ref{fig:fig3}--\ref{fig:fig6}.}
    \label{fig:fig0}
\end{figure}

Our main findings may be summarized as follows. We find that the reconstruction of peaked domain-wall signals at LISA is governed primarily by the position of the spectral turnover relative to the detector sensitivity window. When the peak lies well outside the LISA band, the detector probes only the ultraviolet or infrared side of the spectrum, and the posterior develops strong degeneracies because the signal depends mainly on an effective parameter combination. By contrast, when the peak lies within or near the most sensitive part of the LISA band, the detector becomes sensitive to both the overall amplitude and the local spectral curvature, leading to a substantial reduction of the posterior degeneracy. In this intermediate regime, and provided the signal is sufficiently strong, the marginalized uncertainties in the original signal parameters also become small, enabling genuine two-parameter reconstruction. For the benchmark signal template considered here, the most favorable reconstruction region typically corresponds to annihilation temperatures in the range $10^3\text{--}10^6\,\mathrm{GeV}$, with the most robust reconstruction achieved at ${\rm SNR}\gtrsim 50$. Signals with ${\rm SNR}\sim 10$
can still be reconstructed effectively only in a narrower
part of parameter space concentrated near
$T_*\sim 10^{5-6}\,\mathrm{GeV}$.
In terms of the observable spectrum, this weaker-signal region corresponds approximately to peak amplitudes in the range $4\times10^{-12}\lesssim \Omega_{\rm GW}^{\rm peak}h^2 \lesssim 4\times10^{-11}$, with characteristic peak frequencies satisfying $30\,{\rm mHz}\lesssim f_p\lesssim 100\,{\rm mHz}$, with the precise upper limit depending weakly on the relativistic degrees of freedom. The present work, therefore, identifies the part of the domain-wall parameter space in which LISA has the strongest reconstruction power for peaked signals. A representative illustration of this result is provided by the heat map of $\sigma(\log_{10}T_*)$ in Fig.~\ref{fig:fig0}, which directly shows the region where the annihilation temperature, and equivalently the peak frequency, can be reconstructed most accurately.

The rest of this paper is organized as follows. In Sec.~\ref{sec:dw_gw} we summarize the domain-wall GW signal model and its relevant infrared and ultraviolet limits. In Sec.~\ref{sec:inference_setup} we describe the LISA setup, astrophysical foregrounds, Bayesian likelihood, and posterior-geometry diagnostics. The principal reconstruction results are presented in Sec.~\ref{sec:results}. We conclude in Sec.~\ref{sec:conclusion}.

\section{Domain-wall gravitational-wave signal model}
\label{sec:dw_gw}

Domain walls are two-dimensional topological defects that may form after the spontaneous breaking of a discrete symmetry, for example, a $Z_2$ symmetry \cite{Zeldovich:1974uw,Press:1989yh,Saikawa:2017hiv,Jaeckel:2016jlh,Jiang:2022svq}. After formation, the network rapidly approaches a scaling regime in which the so-called area parameter $\mathcal{A}$ becomes approximately constant and of order unity, as found in lattice simulations \cite{Press:1989yh,Hiramatsu:2013qaa,Dankovsky:2024zvs,Babichev:2025stm,Notari:2025kqq,Cyr:2025nzf,Kitajima:2023cek,Blasi:2025tmn}. In this regime, the wall energy density is well approximated by
\begin{equation}
    \rho_{\rm DW}\simeq \mathcal{A} H \sigma,
\end{equation}
where $H$ is the Hubble parameter and $\sigma$ is the domain-wall tension.

Since the total background energy density scales as $H^2$, the fractional contribution of domain walls to the cosmic energy budget grows as $H^{-1}$. If the walls were perfectly stable, they would therefore eventually dominate the energy density of the Universe, in conflict with standard cosmology. A standard way to avoid this outcome is to introduce a small bias term in the potential, which lifts the degeneracy between the vacua and triggers wall annihilation at a characteristic temperature $T_*$. The anisotropic stress of the wall network then sources a stochastic gravitational-wave background, with the dominant GW production occurring near the annihilation epoch and the characteristic emission scale set by the Hubble scale at that time.

In this work, we adopt the phenomenological present-day GW template used in recent PTA analyses of domain walls \cite{ng5},
\begin{equation}
    \Omega_{\rm GW} h^2
    =
    \frac{3}{32\pi}\,
    \mathcal{D}\,
    \tilde{\epsilon}\,
    \alpha_*^2\,
    \mathcal{S}(f/f_p),
    \label{eq:omega_dw}
\end{equation}
where $\mathcal{D}\simeq 2\times 10^{-5}$ is a dilution factor, $\tilde{\epsilon}\simeq 0.7$ is an efficiency factor \cite{Hiramatsu:2013qaa}, and $\alpha_*$ denotes the fractional energy density stored in domain walls at the annihilation temperature $T_*$. Recent simulations have reported somewhat smaller values, $\tilde{\epsilon}\simeq 0.2\text{--}0.5$ \cite{Babichev:2025stm,Dankovsky:2025pjg,Kitajima:2023kzu}; adopting such values would lower the signal amplitude and therefore lead to a modest degradation of the reconstruction prospects discussed here. The spectral-shape function is modeled as
\begin{equation}
    \mathcal{S}(x)=
    \frac{(a+b)^c}
    {\left(b\,x^{-a/c}+a\,x^{b/c}\right)^c},
    \qquad x \equiv \frac{f}{f_p},
    \label{eq:dw_shape}
\end{equation}
which provides a smooth interpolation between an infrared rise, a turnover near the peak frequency, and an ultraviolet tail. The redshifted peak frequency is given by
\begin{equation}
    f_p =
    1.14\,{\rm mHz}\,
    \left(\frac{10.75}{g_{*,s}}\right)^{1/3}
    \left(\frac{g_*}{10.75}\right)^{1/2}
    \left(\frac{T_*}{10^{4}\,{\rm GeV}}\right),
    \label{eq:fp_dw}
\end{equation}
where $g_*$ and $g_{*,s}$ denote the effective relativistic degrees of freedom in the energy and entropy densities at the annihilation temperature $T_*$.

Throughout the main text, we adopt the benchmark choice
\begin{equation}
    a=3,\qquad b=1,\qquad c=1,
    \label{bmark}
\end{equation}
which yields the familiar broken-power-law behavior
\[
\Omega_{\rm GW}\propto f^3
\qquad (f\ll f_p),
\]
and
\[
\Omega_{\rm GW}\propto f^{-1}
\qquad (f\gg f_p).
\]
As shown in our recent study of PTA-compatible domain walls producing an ultraviolet tail in the LISA band, the reconstruction quality is largely insensitive to moderate variations in the spectral-shape parameters $a$, $b$, and $c$ \cite{Datta:2026fav}. The same conclusion applies in the present peak-in-band analysis. One may therefore, if desired, repeat the analysis using simulation-motivated values of these parameters that differ slightly from the benchmark choice in Eq.~\eqref{bmark}, without qualitatively altering the main reconstruction results.

Since the present analysis is concerned with peaked signals in the LISA band, it is useful to make both asymptotic regimes explicit. In the infrared limit,
\begin{equation}
    f \ll f_p,
\end{equation}
Eq.~\eqref{eq:dw_shape} reduces to
\begin{equation}
    \mathcal{S}(f/f_p)\simeq
    \frac{(a+b)^c}{b^c}
    \left(\frac{f}{f_p}\right)^a.
\end{equation}
For the benchmark choice $a=3$, $b=1$, $c=1$, this becomes
\begin{equation}
    \mathcal{S}(f/f_p)\simeq 4\left(\frac{f}{f_p}\right)^3.
\end{equation}
Substituting Eq.~\eqref{eq:fp_dw} into Eq.~\eqref{eq:omega_dw}, one obtains
\begin{equation}
    \Omega_{\rm GW} h^2
    \propto
    \alpha_*^2 f_p^{-3}
    \propto
    \alpha_*^2 T_*^{-3},
    \qquad f\ll f_p,
    \label{eq:ir_scaling}
\end{equation}
up to factors depending weakly on the thermal history through $g_*$ and $g_{*,s}$, and apart from the explicit frequency dependence $\propto f^3$.

In the ultraviolet limit,
\begin{equation}
    f \gg f_p,
\end{equation}
Eq.~\eqref{eq:dw_shape} simplifies instead to
\begin{equation}
    \mathcal{S}(f/f_p)\simeq
    \frac{(a+b)^c}{a^c}
    \left(\frac{f}{f_p}\right)^{-b}.
\end{equation}
For the benchmark choice, this gives
\begin{equation}
    \mathcal{S}(f/f_p)\simeq \frac{4}{3}\left(\frac{f}{f_p}\right)^{-1},
\end{equation}
and therefore
\begin{equation}
    \Omega_{\rm GW} h^2
    \propto
    \alpha_*^2 f_p
    \propto
    \alpha_*^2 T_*,
    \qquad f\gg f_p,
    \label{eq:uv_scaling}
\end{equation}
again up to weakly varying thermal-history factors and the explicit frequency dependence $\propto f^{-1}$.

These limiting behaviors clarify the reconstruction problem in different parts of parameter space. If the peak frequency lies well below the detector band, the observed signal is dominated by the ultraviolet tail and depends mainly on the effective combination $\alpha_*^2 T_*$. In this regime, many parameter pairs satisfying approximately
\begin{equation}
    \alpha_*^2 T_* \simeq \mathrm{const.}
\end{equation}
produce nearly indistinguishable in-band spectra, leading to the familiar strong anti-correlated degeneracy in the posterior \cite{Datta:2026fav}. On the other hand, if the peak lies well above the detector band, the detector probes mainly the infrared side of the spectrum, and the signal scales approximately as
\begin{equation}
    \alpha_*^2 T_*^{-3} \simeq \mathrm{const.},
\end{equation}
which again implies an approximate degeneracy, now with a different orientation in parameter space. In logarithmic variables, these two asymptotic relations become
\begin{equation}
    2\log_{10}\alpha_* + \log_{10}T_* \simeq \mathrm{const.},
    \qquad (f\gg f_p),
\end{equation}
and
\begin{equation}
    2\log_{10}\alpha_* - 3\log_{10}T_* \simeq \mathrm{const.},
    \qquad (f\ll f_p).
\end{equation}

The intermediate regime, in which the peak lies within or near the LISA band, is qualitatively different. There, the detector becomes sensitive not only to the overall signal amplitude but also to the spectral turnover itself, so that the two signal parameters affect the observed spectrum in more distinct ways. One therefore expects the posterior degeneracy to be substantially reduced and, in the most favorable part of parameter space, genuine two-parameter reconstruction to become possible. This transition from infrared- or ultraviolet-tail-dominated inference to peak-dominated inference is the central theme of the reconstruction analysis presented below.

\section{Inference setup for LISA}
\label{sec:inference_setup}

In this section, we briefly summarize the detector model, astrophysical foreground treatment, Bayesian likelihood, and posterior-geometry diagnostics used in the analysis; an expanded discussion is given in Ref.~\cite{Datta:2026fav}. Since our aim is to map where peaked domain-wall signals can be reconstructed most effectively, we retain both the instrumental signal-to-noise ratio (SNR) as a basic detectability measure and the posterior covariance structure as a diagnostic of parameter identifiability.

\subsection{TDI channels, detector response, and instrumental noise}

The LISA data are described in terms of the orthogonal time-delay interferometry (TDI) channels $A$, $E$, and $T$, obtained from suitable linear combinations of the raw unequal-arm interferometric measurements \cite{Tinto:2004wu,McNamara:2008zz,Armano:2018kix,Caprini:2019pxz,Flauger:2020qyi,Caprini:2024hue}. In the equal-arm approximation, the one-sided PSD in channel $a\in\{A,E,T\}$ is written as
\begin{equation}
    P_a(f)=S_a(f)+N_a(f),
\end{equation}
where $S_a(f)$ is the stochastic GW signal contribution and $N_a(f)$ is the instrumental-noise contribution.

The signal contribution is related to the GW energy-density spectrum by
\begin{equation}
    S_a(f)
    =
    \frac{3H_0^2}{4\pi^2}\,
    \frac{\Omega_{\rm GW}(f)}{f^3}\,
    \mathcal R_a(f),
\end{equation}
where $\mathcal R_a(f)$ denotes the channel response function. For LISA, we take
\begin{equation}
    L = 2.5\times 10^9~{\rm m},
    \qquad
    \delta x = 15\times 10^{-12}~{\rm m}/\sqrt{\rm Hz},
\end{equation}
with transfer frequency
\begin{equation}
    f_\star=\frac{c}{2\pi L}.
\end{equation}
The response functions are modeled in the standard form \cite{Cornish:2001bb,Smith:2019wny}
\begin{equation}
    \mathcal R_A(f)=\mathcal R_E(f)
    =
    \frac{9}{5}|W(f)|^2
    \left[
    1+\left(\frac{f}{4f_\star/3}\right)^2
    \right]^{-1},
\end{equation}
\begin{equation}
    \mathcal R_T(f)
    =
    \frac{1}{1008}|W(f)|^2
    \left(\frac{f}{f_\star}\right)^6
    \left[
    1+\frac{5}{16128}\left(\frac{f}{f_\star}\right)^8
    \right]^{-1},
\end{equation}
with
\begin{equation}
    W(f)=1-e^{-2if/f_\star}.
\end{equation}

The instrumental noise is described by the standard acceleration and optical-metrology contributions,
\begin{equation}
    \sqrt{S_{\rm acc}(f)}
    =
    N_{\rm acc}
    \sqrt{1+\left(\frac{0.4~{\rm mHz}}{f}\right)^2}
    \sqrt{1+\left(\frac{f}{8~{\rm mHz}}\right)^4},
\end{equation}
\begin{equation}
    \sqrt{S_{\rm OMS}(f)}
    =
    \delta x
    \sqrt{1+\left(\frac{2~{\rm mHz}}{f}\right)^4},
\end{equation}
with $N_{\rm acc}=3\times 10^{-15}~{\rm m\,s^{-2}/\sqrt{Hz}}$. These components are combined in the usual equal-arm TDI expressions to obtain the channel noise PSDs $N_A(f)$, $N_E(f)$, and $N_T(f)$. At low frequencies, the $T$ channel has a strongly suppressed GW response and therefore acts approximately as a null channel.

For later use, we define the effective channel sensitivity
\begin{equation}
    \mathcal S_j(f)=\sqrt{\frac{N_j(f)}{\mathcal R_j(f)}},
    \qquad j\in\{A,E,T\},
\end{equation}
and the channel-by-channel signal-to-noise ratio
\begin{equation}
    {\rm SNR}_j
    =
    \left[
    2T_t\int_0^\infty
    \left(\frac{S_j(f)}{N_j(f)}\right)^2
    df
    \right]^{1/2},
    \qquad
    j\in\{A,E,T\},
\end{equation}
where $T_t$ is the total observation time. In practice, the integral is restricted to the analyzed LISA band. The $A$ and $E$ channels carry the dominant signal sensitivity, while the $T$ channel serves primarily as a noise monitor.

\subsection{Astrophysical foregrounds}

A cosmological stochastic background in the LISA band is observed in the presence of astrophysical foregrounds. The dominant contribution is expected from unresolved Galactic double white-dwarf (DWD) binaries \cite{Korol:2021pun,Korol:2020lpq,Liu:2023qap}, whose superposition forms the familiar confusion foreground in the milli-Hz range. In addition, unresolved extra-galactic compact binaries generate a smoother stochastic background \cite{Phinney:2001di,Regimbau:2011rp,Babak:2023lro,Lehoucq:2023zlt}. We therefore model the total GW spectrum as
\begin{equation}
    \Omega_{\rm GW}(f)
    =
    \Omega_{\rm dwd}(f)
    +
    \Omega_{\rm GW,ast}(f)
    +
    \Omega_{\rm DW}(f),
\end{equation}
where the three terms denote, respectively, the Galactic DWD foreground, the extra-galactic astrophysical background, and the domain-wall signal of interest.

For the Galactic foreground, we adopt the phenomenological broken-power-law form
\begin{equation}
    \Omega_{\rm dwd}(f)
    =
    \frac{A_1 (f/f_\star)^{\alpha_1}}
    {1+A_2 (f/f_\star)^{\alpha_2}},\label{eq:galac}
\end{equation}
which captures the broad rise and turnover of the unresolved DWD confusion signal \cite{Korol:2021pun,Korol:2020lpq,Liu:2023qap,Chen:2023zkb}. The extra-galactic astrophysical background is modeled as a single power law,
\begin{equation}
    \Omega_{\rm GW,ast}(f)
    =
    \Omega_{\rm ast}
    \left(\frac{f}{f_\star}\right)^{\varepsilon},\label{eq:exgalac}
\end{equation}
following standard stochastic-background forecasts \cite{Phinney:2001di,Regimbau:2011rp,Babak:2023lro,Lehoucq:2023zlt,Boileau:2020rpg}. In the present work, we use isotropic effective templates for both components and neglect anisotropy and annual modulation of the Galactic foreground \cite{Giampieri:1997ie,Criswell:2024hfn,Digman:2022jmp,Hindmarsh:2024ttn,Buscicchio:2024wwm}.

The full set of free parameters is listed in Table~\ref{tab:model_parameters}. These include the two domain-wall signal parameters together with eight nuisance parameters describing instrumental noise and astrophysical foregrounds.

\begin{table}[htbp]
\centering
\begin{tblr}
	{
    hlines,
    vlines,
    row{1} = {bg=gray7, fg=white, font=\bfseries},
column{1} = {bg=gray9},
cell{1}{1} = {bg=gray7, fg=white},
}

\textbf{Parameter} & \textbf{Injected value} & \textbf{Prior (uniform)} \\
$\log_{10}(N_{\rm acc})$      & $-14.523$ & $(-16.0,-13.7)$ \\
$\log_{10}(\delta_x)$         & $-11.097$ & $(-13.0,-10.7)$ \\
$\log_{10}(A_1)$              & $-15.4$   & $(-18.0,-5.0)$ \\
$\alpha_1$                    & $-5.7$    & $(-15.0,-3.0)$ \\
$\log_{10}(A_2)$              & $-6.32$   & $(-10.0,5.0)$ \\
$\alpha_2$                    & $-6.2$    & $(-10.0,-1.0)$ \\
$\log_{10}(\Omega_{\rm ast})$ & $-11.0$   & $(-15.0,-8.0)$ \\
$\varepsilon$                 & $0.67$    & $(0.0,1.0)$ \\
$\log_{10}(T_\star/\rm GeV)$          & grid      & $(1.74,7.15)$ \\
$\log_{10}(\alpha_\star)$     & grid      & $(-5.0,0.5)$ \\
\end{tblr}
\caption{Model parameters, fiducial values, and uniform prior ranges used in the inference.}
\label{tab:model_parameters}
\end{table}

\subsection{Frequency-domain likelihood and mock data generation}

The Bayesian analysis is performed in the frequency domain. Let $\tilde d_a^\kappa(f_k)$ denote the discrete Fourier transform of the $\kappa$-th time segment in channel $a\in\{A,E,T\}$ at frequency bin $f_k$, and define the data vector
\begin{equation}
    \mathbf d_{\kappa k}
    =
    \begin{pmatrix}
        \tilde d_A^\kappa(f_k)\\
        \tilde d_E^\kappa(f_k)\\
        \tilde d_T^\kappa(f_k)
    \end{pmatrix}.
\end{equation}
For each segment and frequency bin, we model $\mathbf d_{\kappa k}$ as a zero-mean complex Gaussian variable with diagonal covariance
\begin{equation}
    C_k(\theta)
    =
    \frac{T f_s^2}{2}
    \begin{pmatrix}
        P_A(f_k;\theta) & 0 & 0\\
        0 & P_E(f_k;\theta) & 0\\
        0 & 0 & P_T(f_k;\theta)
    \end{pmatrix},
\end{equation}
where $T$ is the segment duration, $f_s$ is the sampling frequency, and $\theta$ denotes the full parameter vector. In the null-channel approximation,
\begin{equation}
    P_A=S_A+N_A,\qquad
    P_E=S_E+N_E,\qquad
    P_T=N_T.
\end{equation}

Assuming independence between different segments and frequency bins, the likelihood takes the form
\begin{equation}
    \mathcal L(\theta)
    =
    \prod_{\kappa,k}
    \frac{1}{\pi^3 \det C_k(\theta)}
    \exp\!\left[-\mathbf d_{\kappa k}^\dagger C_k^{-1}(\theta)\mathbf d_{\kappa k}\right].
\end{equation}
Dropping additive constants independent of $\theta$, the log-likelihood becomes
\begin{equation}
\begin{aligned}
    \ln \mathcal L(\theta)
    =
    -\sum_{\kappa,k}
    \Bigg[
        &\ln\!\Big(P_A(f_k;\theta)\,P_E(f_k;\theta)\,P_T(f_k;\theta)\Big) \\
        &+
        \frac{2}{T f_s^2}
        \left(
        \frac{|\tilde d_A^\kappa(f_k)|^2}{P_A(f_k;\theta)}
        +
        \frac{|\tilde d_E^\kappa(f_k)|^2}{P_E(f_k;\theta)}
        +
        \frac{|\tilde d_T^\kappa(f_k)|^2}{P_T(f_k;\theta)}
        \right)
    \Bigg].
\end{aligned}
\label{eq:loglike_compact}
\end{equation}
This is the likelihood used throughout the numerical analysis. A Gaussian-residual likelihood, sometimes employed in the literature \cite{Caprini:2019pxz,Flauger:2020qyi,Samanta:2025jec}, becomes effectively equivalent to the above form in the limit of many averaged segments.

Mock datasets are generated from a fiducial injection model $\theta_{\rm inj}$ by drawing
\begin{equation}
    \mathbf d_{\kappa k}^{\rm mock}
    \sim
    \mathcal N_{\mathbb C}\!\left(0,C_k^{\rm inj}\right).
\end{equation}
The posterior distribution is then
\begin{equation}
    p(\theta|d)
    =
    \frac{\mathcal L(d|\theta)\,\pi(\theta)}{Z},
\end{equation}
where $\pi(\theta)$ denotes the prior and $Z$ the Bayesian evidence. Parameter-space exploration is performed using the \texttt{dynesty} sampler \cite{speagle2020dynesty}, interfaced through the \texttt{Bilby} package \cite{Ashton:2018jfp}.

\subsection{Posterior covariance and reconstruction metrics}

To quantify how well the signal parameters can be reconstructed, we characterize the posterior geometry through the sample covariance matrix
\[
C={\rm Cov}(\boldsymbol{\theta}),
\qquad
\boldsymbol{\theta}=(\theta_1,\theta_2,\dots,\theta_N)^T.
\]
Its eigendecomposition
\[
C=E\Lambda E^T,
\qquad
\Lambda={\rm diag}(\lambda_1,\lambda_2,\dots,\lambda_N),
\]
defines the principal directions and the corresponding variances of the posterior cloud. In the two-dimensional signal sector,
\[
\theta_1=\log_{10}\alpha_*,
\qquad
\theta_2=\log_{10}T_*,
\]
we denote the larger and smaller eigenvalues by
\[
\lambda_\parallel,\qquad \lambda_\perp,
\]
corresponding respectively to the long and short axes of the posterior.

The ratio
\begin{equation}
    \kappa \equiv \frac{\lambda_\parallel}{\lambda_\perp}
\end{equation}
provides a direct measure of posterior degeneracy: $\kappa\gg 1$ corresponds to a strongly elongated posterior, whereas $\kappa\sim 1$ indicates a nearly isotropic one. The quantity $\lambda_\perp$ is particularly useful because it measures the width of the genuinely constrained parameter combination, while the broad direction associated with $\lambda_\parallel$ can be more sensitive to prior volume. By contrast, the area of a constant-density ellipse scales as
\[
A\propto \sqrt{\lambda_\parallel\lambda_\perp},
\]
so area-based summaries are automatically contaminated by the broad direction.

Because the signal parameters are defined logarithmically, the corresponding $1\sigma$ widths along the principal axes are $\sqrt{\lambda_\parallel}$ and $\sqrt{\lambda_\perp}$ in dex. At the same time, we also compute the marginalized standard deviations in the original signal coordinates,
\[
\sigma_{\log_{10}\alpha_*}=\sqrt{C_{11}},
\qquad
\sigma_{\log_{10}T_*}=\sqrt{C_{22}}.
\]
The covariance matrix also defines the correlation coefficient
\[
\rho =\frac{C_{12}}{\sqrt{C_{11}C_{22}}},
\]
which measures the degree of degeneracy between $\log_{10}{\alpha_*}$ and $\log_{10}{T_*}$.
These quantities are useful because in the low-degeneracy regime, where $\kappa$ approaches unity, the marginalized uncertainties and the principal-component widths become comparable. Comparing the heat maps of $\sigma_{\log_{10}\alpha_*}$ and $\sigma_{\log_{10}T_*}$ with those of $\lambda_\parallel$ and $\lambda_\perp$ therefore provides a direct way to distinguish genuinely good two-parameter reconstruction from a merely weakly elongated posterior.

A convenient reconstruction threshold is obtained by requiring the positive-side fractional uncertainty along the perpendicular direction to remain below $25\%$, which gives \cite{Datta:2026fav}
\begin{equation}
    \log_{10}\!\big(\sqrt{\lambda_\perp}\big)\lesssim -1.01.
\end{equation}
Values below this threshold therefore indicate better than $25\%$ reconstruction along the constrained parameter combination.

The covariance matrix also allows uncertainties to be propagated from parameter space to signal space. If the model signal is $S(f;\boldsymbol{\theta})$, the corresponding local signal variance is approximated by
\begin{equation}
\sigma_S^2(f)
\simeq
\nabla_\theta S(f)^T\,C\,\nabla_\theta S(f),
\end{equation}
or, in the principal-component basis,
\begin{equation}
\sigma_S^2(f)
\simeq
\left(\frac{\partial S}{\partial q_\parallel}\right)^2\lambda_\parallel
+
\left(\frac{\partial S}{\partial q_\perp}\right)^2\lambda_\perp.
\end{equation}
In practice, we evaluate this using finite displacements along the perpendicular mode. If
\begin{equation}
\boldsymbol{\theta}_\pm
=
\boldsymbol{\theta}_0 \pm \sqrt{\lambda_\perp}\,\mathbf e_\perp,
\end{equation}
then the corresponding signal uncertainty is approximated by
\begin{equation}
\sigma_S(f)
\simeq
\frac{|S_+(f)-S_-(f)|}{2},
\end{equation}
with fractional uncertainty
\begin{equation}
\frac{\sigma_S(f)}{S(f;\boldsymbol{\theta}_0)}
\simeq
\frac{|S_+(f)-S_-(f)|}{2\,S(f;\boldsymbol{\theta}_0)}.
\end{equation}

These covariance-based diagnostics are central to the analysis below. The SNR maps identify where the signal is detectable, while the combinations of $\kappa$, $\lambda_\perp$, $\lambda_\parallel$, and the marginalized standard deviations $\sigma_{\log_{10}\alpha_*}$ and $\sigma_{\log_{10}T_*}$ determine where the posterior is both tight and weakly degenerate, thereby allowing genuine two-parameter reconstruction.
\section{Parameter reconstruction landscape in the LISA band}
\label{sec:results}

We now turn to the main reconstruction results for peaked domain-wall signals in the LISA band. The central objective is to identify where in the $(\alpha_*,T_*)$ plane the signal can be reconstructed with both small uncertainty and weak parameter degeneracy. As discussed in the previous sections, these are distinct requirements: a posterior may be nearly isotropic but broad, or highly elongated while still being sharply constrained across one direction. The results below therefore combine several complementary diagnostics, namely the degeneracy measures $\kappa$ and $1-\rho^2$, the marginalized uncertainties $\sigma_{\alpha_*}$ and $\sigma_{T_*}$, and the perpendicular variance $\lambda_\perp$.
\begin{figure}
    \centering
   \includegraphics[width=0.5\linewidth]{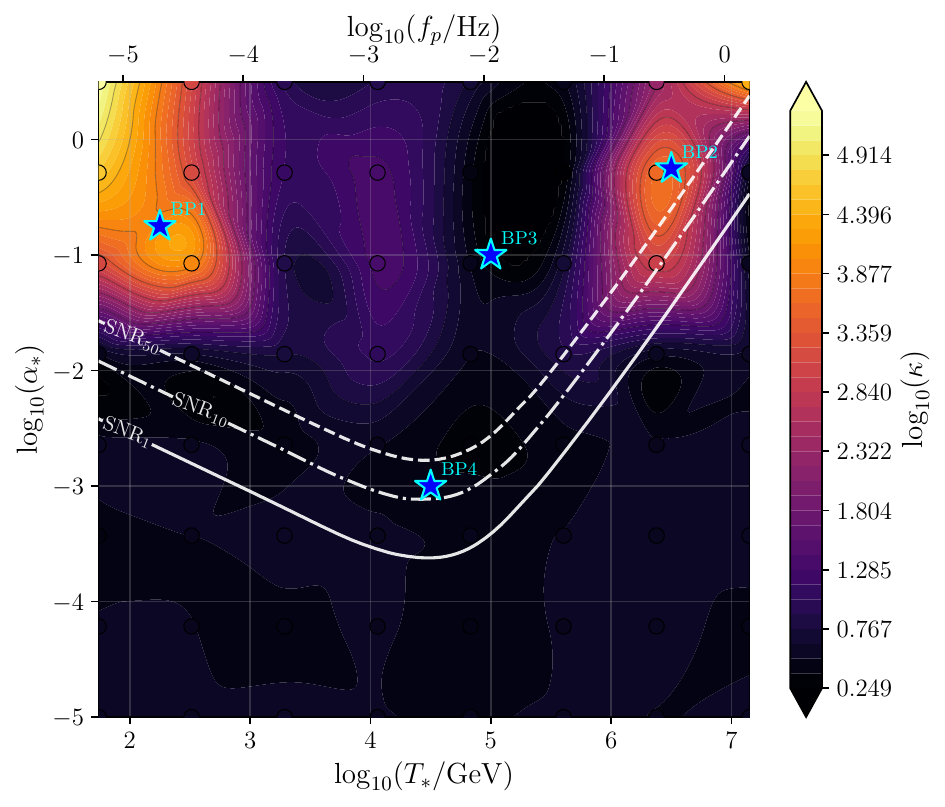}\includegraphics[width=0.5\linewidth]{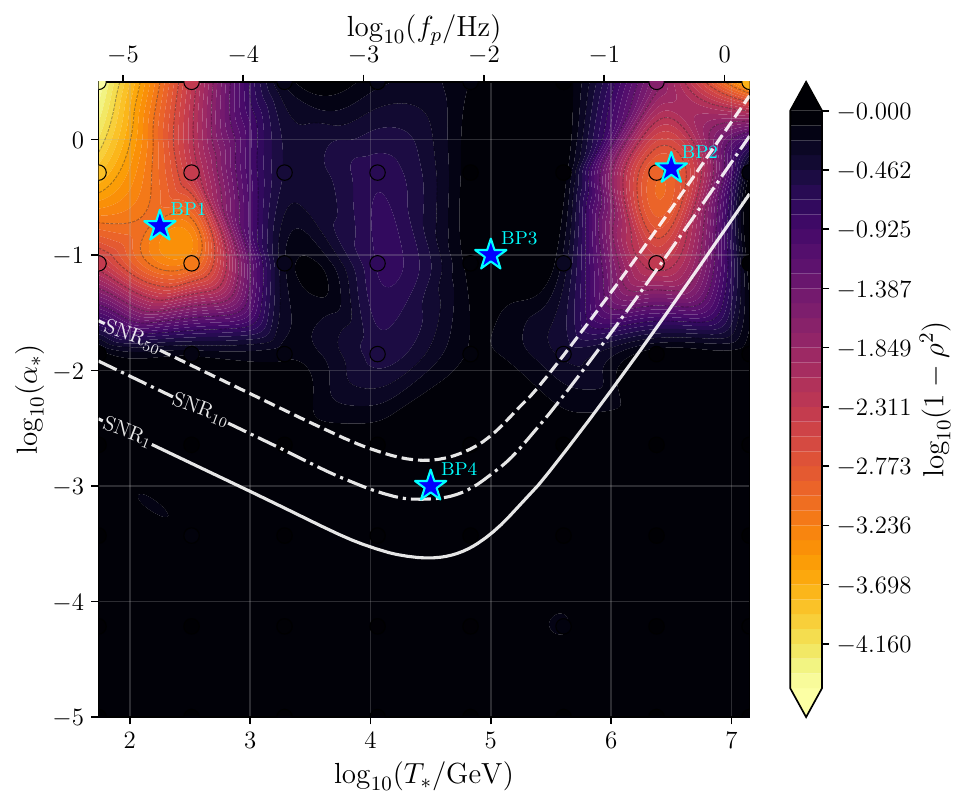}\\
    \includegraphics[width=0.5\linewidth]{figure2d_sigma_T.pdf}\includegraphics[width=0.5\linewidth]{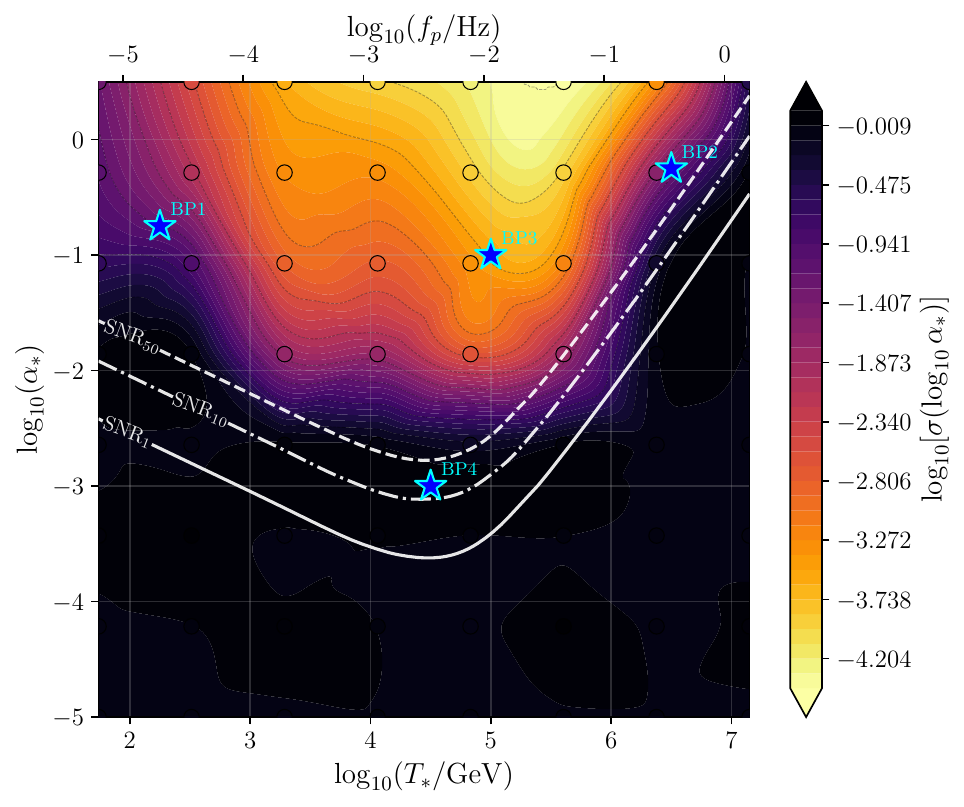}
    \caption{Top left: heat map of $\kappa$ on the $\alpha_*$--$T_*$ plane. The bright region toward the upper left corresponds to a UV-tail-dominated degeneracy, while the bright region toward the upper right corresponds to an IR-tail-dominated degeneracy. The dark central band, where $\kappa\simeq 1$, marks the least degenerate part of parameter space and therefore the region most favorable for genuine two-parameter reconstruction. The white contours show ${\rm SNR}=1,\,10$, and $50$. Regions with ${\rm SNR}\lesssim 10$ should, however, be interpreted with care: there the LISA likelihood is weak, so small $\kappa$ may simply reflect a broad, nearly isotropic posterior rather than a truly informative reconstruction. Top right: heat map of $1-\rho^2$, which provides an independent measure of posterior degeneracy and conveys the same qualitative picture as the $\kappa$ map. Bottom left and bottom right: heat maps of $\sigma_{\alpha_*}$ and $\sigma_{T_*}$, respectively. These broadly follow the same pattern as the degeneracy maps and show that the smallest marginalized uncertainties occur in the same central region where the posterior is least elongated. In the strongly degenerate tail-dominated regions, however, the marginalized widths alone are not the most faithful indicators of reconstruction quality; for that purpose $\lambda_\perp$ is more informative, as shown in Fig.~2. The four-starred benchmark points are used to generate the corner plots in Figs.~\ref{fig:fig3}--\ref{fig:fig6}: BP1 samples the UV-tail-dominated region, BP2 the IR-tail-dominated region, BP3 the minimally degenerate and best-reconstructed region, and BP4 a lower-$\alpha_*$ point with similar $T_*$ to BP3, illustrating a weak-signal case with poor reconstruction prospects.}
    \label{fig:fig1}
\end{figure}
\begin{figure}
    \centering
    \includegraphics[width=0.5\linewidth]{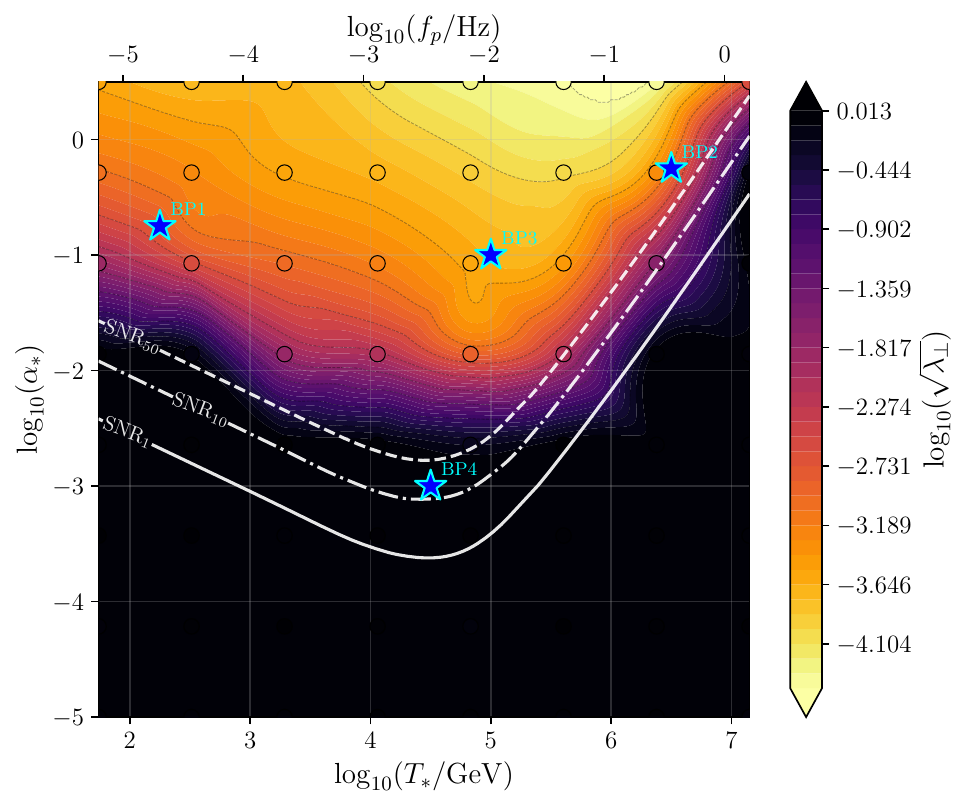}\includegraphics[width=0.5\linewidth]{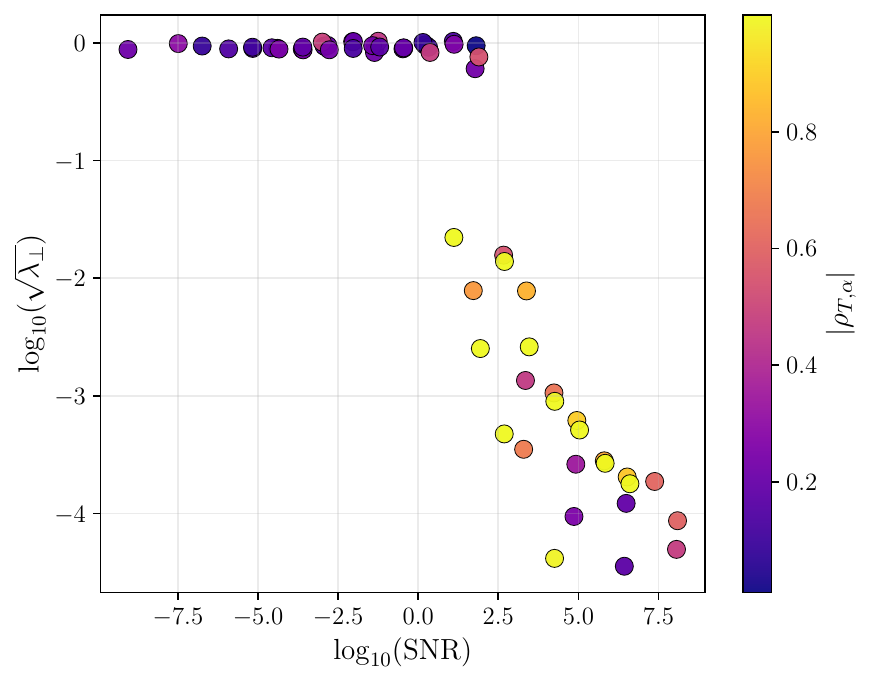}
    \caption{Left: heat map of $\lambda_\perp$ on the $\alpha_*$--$T_*$ plane. Here $\lambda_\perp$ denotes the variance along the short axis of the posterior and therefore quantifies the width of the genuinely constrained parameter combination. Unlike area-based summaries, it is not contaminated by the broad, potentially prior-sensitive direction associated with posterior elongation. Small values of $\lambda_\perp$ thus identify the region of strongest local reconstruction power. The map shows that LISA constrains the signal most efficiently in an intermediate region of parameter space, whereas the reconstruction deteriorates toward the infrared- and ultraviolet-tail-dominated limits. Right: $\lambda_\perp$ as a function of SNR. The plateau at low SNR signals the onset of prior-dominated inference, while the decrease at higher SNR marks the regime in which the LISA likelihood begins to constrain the signal more effectively.}
    \label{fig:fig2}
\end{figure}
The overall reconstruction landscape is summarized in Fig.~\ref{fig:fig1}. The top-left panel shows the heat map of $\kappa$ across the signal-parameter plane. A clear structure is visible. Toward the upper-left part of the plane, the posterior becomes strongly elongated because the signal in the LISA band is dominated by the ultraviolet tail; in that regime, many parameter combinations generate nearly indistinguishable spectra, leading to a large value of $\kappa$. Toward the upper-right part of the plane, a second strongly degenerate region appears, now associated with the infrared tail. In that regime, the detector probes mainly the rising side of the spectrum, and the corresponding approximate scaling relation again produces a broad posterior ridge, but with a different orientation in parameter space. Between these two tail-dominated limits, there is an extended dark band in which $\kappa\simeq 1$, indicating that the posterior is much less elongated and that the two signal parameters can, in principle, be reconstructed simultaneously.

\begin{figure}
    \centering
    \includegraphics[width=1.1\linewidth]{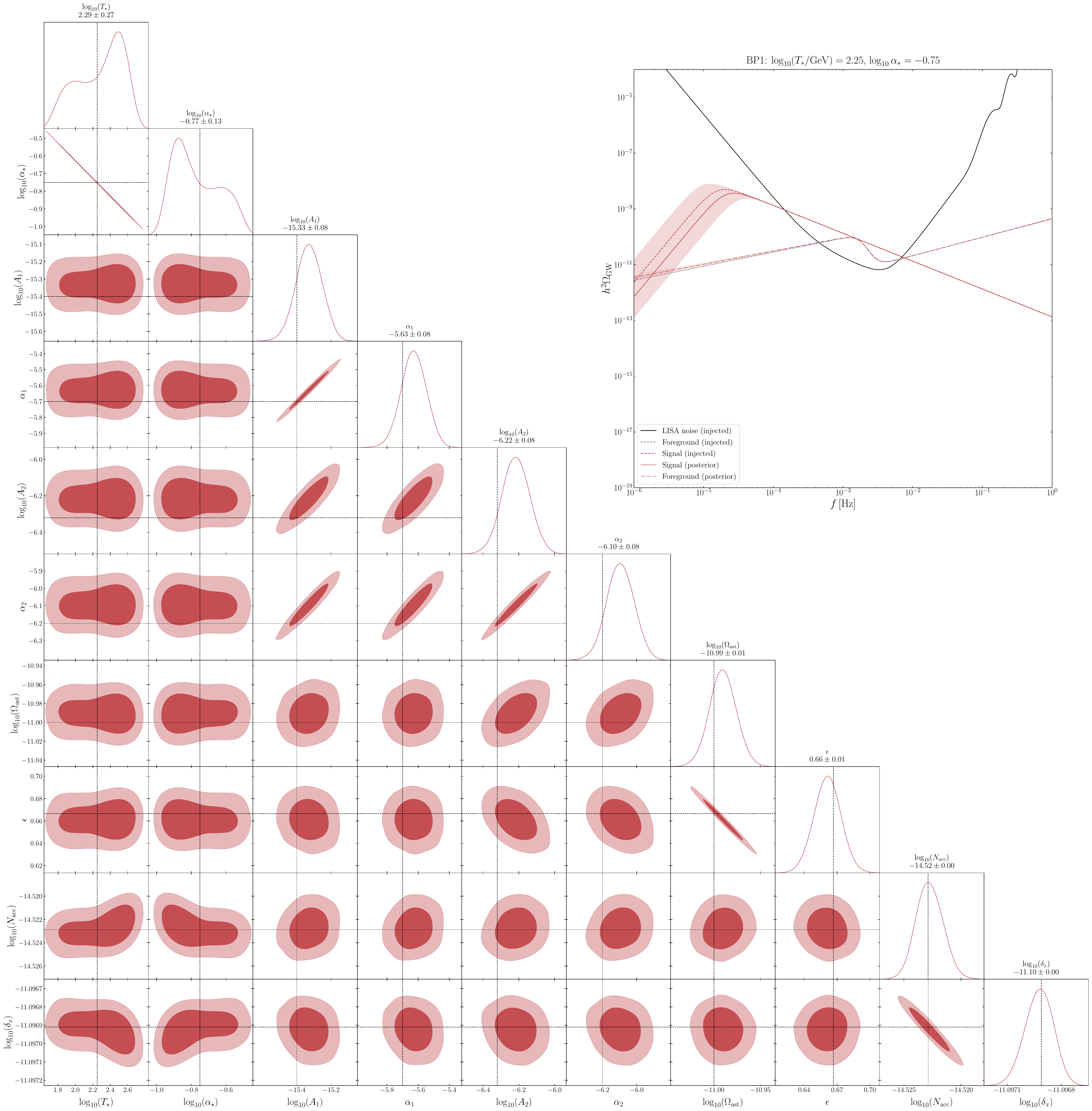}
    \caption{Corner plot for BP1, shown in Fig.~\ref{fig:fig1}, illustrating the posterior structure in the UV-tail-dominated region. The top-right inset displays the corresponding signal reconstruction on the $\Omega_{\rm GW}h^2$--$f$ plane.}
    \label{fig:fig3}
\end{figure}
\begin{figure}
    \centering
    \includegraphics[width=1.1\linewidth]{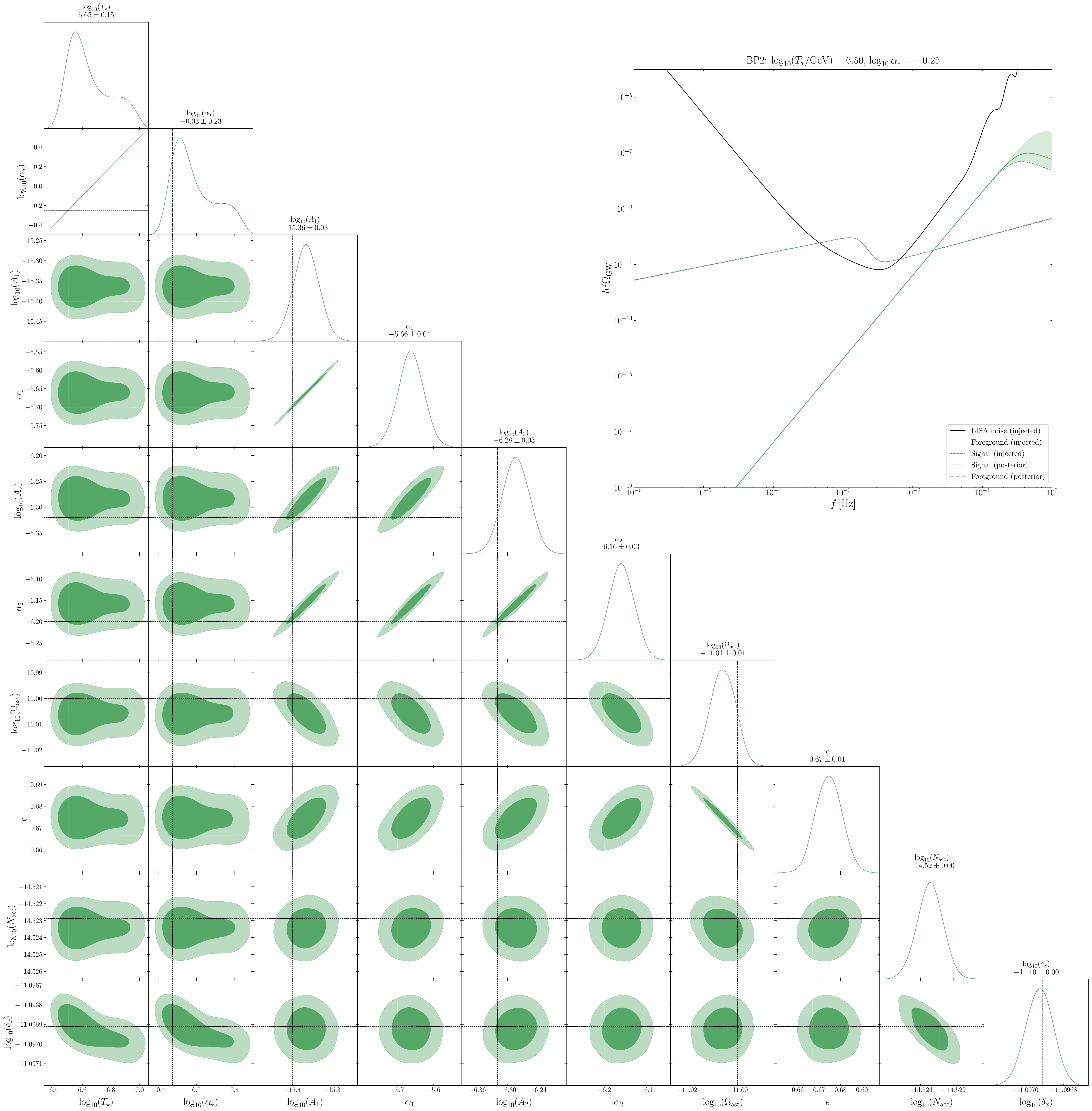}
  \caption{Corner plot for BP2, shown in Fig.~\ref{fig:fig1}, illustrating the posterior structure in the IR-tail-dominated region. The top-right inset displays the corresponding signal reconstruction on the $\Omega_{\rm GW}h^2$--$f$ plane.}
    \label{fig:fig4}
\end{figure}
\begin{figure}
    \centering
    \includegraphics[width=1.1\linewidth]{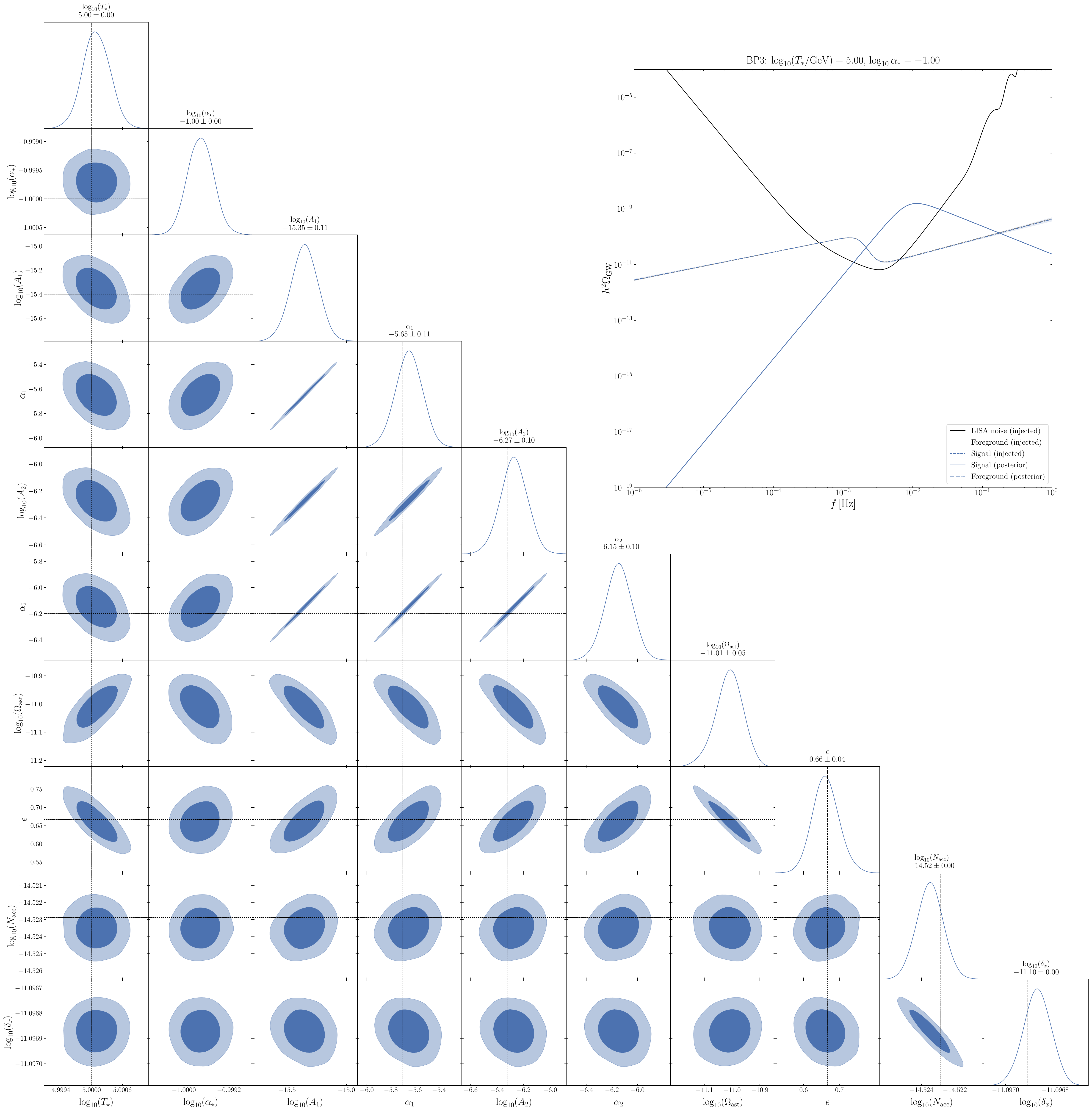}
    \caption{Corner plot for BP3, shown in Fig.~\ref{fig:fig1}, illustrating the posterior structure in the minimally degenerate region, where the spectral turnover lies within or near the most sensitive part of the LISA band. The top-right inset displays the corresponding signal reconstruction on the $\Omega_{\rm GW}h^2$--$f$ plane.}
    \label{fig:fig5}
\end{figure}
\begin{figure}
    \centering
    \includegraphics[width=1.1\linewidth]{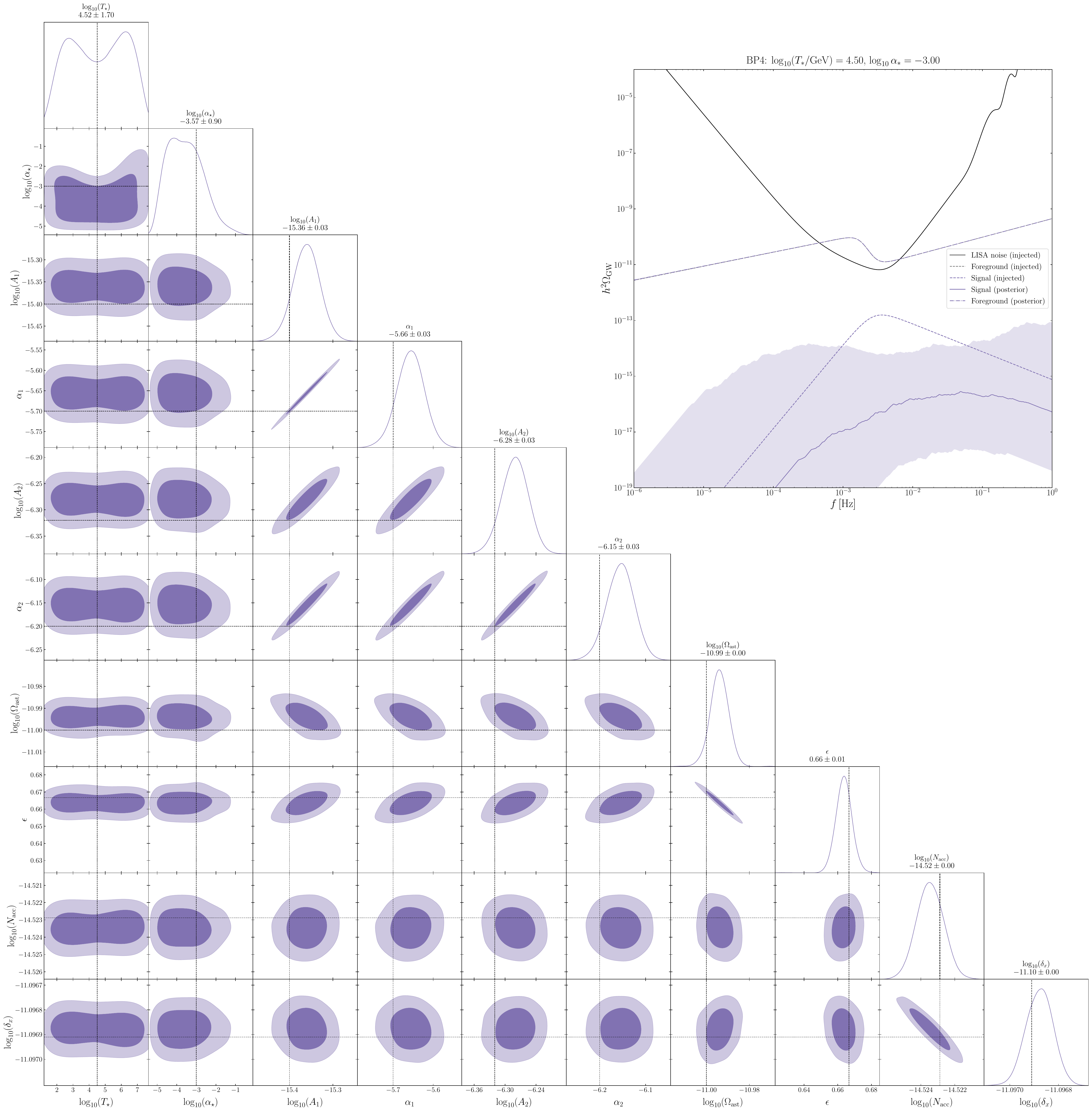}
    \caption{Corner plot for BP4, shown in Fig.~\ref{fig:fig1}, illustrating the posterior structure for a weak-signal benchmark with poor reconstruction prospects despite a similar peak location to BP3. The top-right inset displays the corresponding signal reconstruction on the $\Omega_{\rm GW}h^2$--$f$ plane.}
    \label{fig:fig6}
\end{figure}
This interpretation is reinforced by the top-right panel, which shows the independent degeneracy diagnostic $1-\rho^2$. The pattern closely mirrors that of the $\kappa$ map: the tail-dominated regions correspond to strongly correlated posteriors, whereas the central band is much less correlated. The agreement between the two panels shows that the reduced degeneracy in the middle of the plane is not an artifact of a particular metric, but a genuine feature of the posterior geometry.

The white SNR contours overlaid on the top-left panel are essential for interpreting this structure. In particular, the central low-$\kappa$ region should not be read naively as uniformly favorable for reconstruction. At low signal strength, especially below ${\rm SNR}\sim 10$, the LISA likelihood becomes weak, and the posterior may become broad but nearly isotropic. In that regime, $\kappa$ alone is not a reliable measure of informative reconstruction: a small value can simply reflect a prior-dominated posterior with comparable widths in both directions. For this reason, the most physically meaningful low-degeneracy region is the part of the central band that also lies above the intermediate SNR contours. This already suggests one of the main messages of the paper: good reconstruction requires not only weak degeneracy, but also sufficient signal strength.

The bottom panels of Fig.~\ref{fig:fig1} show the marginalized uncertainties $\sigma_{\alpha_*}$ and $\sigma_{T_*}$. These broadly track the same structure as the degeneracy maps. In the central region, where the peak lies within or near the most sensitive part of the detector band, both marginalized uncertainties become small, indicating that the posterior is not only less elongated but also genuinely tighter. By contrast, in the two-tailed-dominated regions, the marginalized uncertainties increase, reflecting the fact that one parameter combination remains poorly determined. The important point is that the best reconstruction region is singled out simultaneously by all four diagnostics: the posterior is least degenerate there, and both signal parameters are constrained most tightly there.

Figure~\ref{fig:fig2} complements this picture by showing the heat map of $\lambda_\perp$, together with its relation to the SNR. As emphasized earlier, $\lambda_\perp$ is the most faithful single quantity for characterizing the local constraining power of the data, since it isolates the short axis of the posterior and is insensitive to the broad, potentially prior-dominated direction. The map confirms that the central low-degeneracy band is also the region in which the perpendicular mode is most sharply constrained. In the low-SNR regime, however, $\lambda_\perp$ approaches a plateau, signaling the onset of prior-dominated inference. This explains why some regions with $\kappa\simeq 1$ do not in fact correspond to high-quality reconstruction: the posterior may be only weakly elongated simply because the likelihood is too weak to constrain either direction strongly.

Taken together, these maps show that the reconstruction problem is controlled by the position of the signal peak relative to the detector band. When the peak lies well below the LISA band, the detector sees only the ultraviolet tail, and the posterior develops the familiar UV-tail degeneracy. When the peak lies well above the band, the signal is probed mainly through its infrared side, producing a different but equally strong degeneracy. Only in the intermediate regime, where the turnover itself lies within or near the sensitive part of the band, does the posterior become simultaneously tight and weakly degenerate. This is precisely the regime in which LISA can move from constraining a single effective parameter combination to genuine two-parameter reconstruction.

To make this structure more explicit, we select four representative benchmark points marked by stars in Fig.~\ref{fig:fig1}. BP1 lies in the UV-tail-dominated region and illustrates the high-$\kappa$ degeneracy that arises when only the falling side of the spectrum is observed. BP2 lies in the IR-tail-dominated region and shows the analogous degeneracy associated with the rising side of the spectrum. BP3 is chosen from the central low-$\kappa$, high-SNR region and represents a nice case for true two-parameter reconstruction. BP4 has a similar peak location to BP3 but lower $\alpha_*$, and therefore serves as a weak-signal example in which the nominally favorable geometry is undermined by insufficient likelihood support. The corresponding corner plots, shown in Figs.~\ref{fig:fig3}--\ref{fig:fig6}, confirm this interpretation directly at the posterior level.

The reconstruction maps, therefore, lead to a simple and physically transparent picture. The key question is not merely where the signal is detectable, but where it is both detectable and geometrically identifiable. In the present model, this occurs when the peak lies roughly in the range corresponding to annihilation temperatures of order $10^3$--$10^6\,{\rm GeV}$ and when the signal-to-noise ratio is at least of order ${\rm SNR}\sim 50$. In this regime, the posterior becomes both tighter and less degenerate, and the marginalized uncertainties in $\alpha_*$ and $T_*$ become comparable to the principal-component widths. This is the region of parameter space in which LISA has genuine two-parameter reconstruction power for peaked domain-wall signals.

\subsection*{Additional remarks on the reconstruction maps}

We conclude the reconstruction analysis with a few comments on the interpretation and possible refinement of the maps.

\begin{figure}
    \centering
    \includegraphics[width=0.48\linewidth]{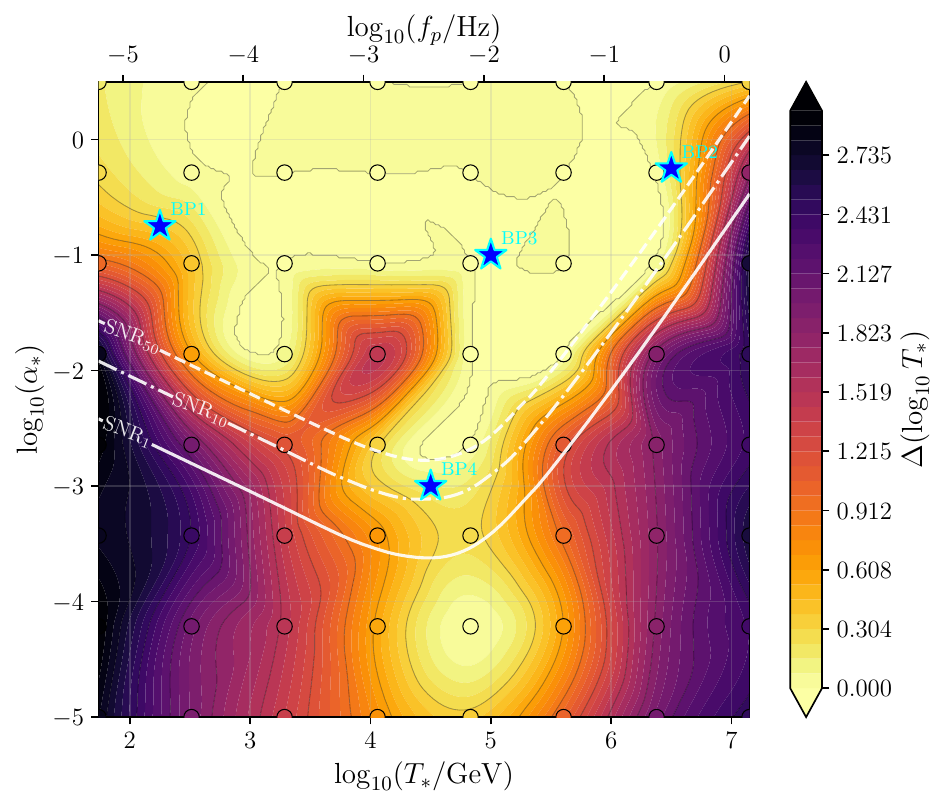}
    \includegraphics[width=0.48\linewidth]{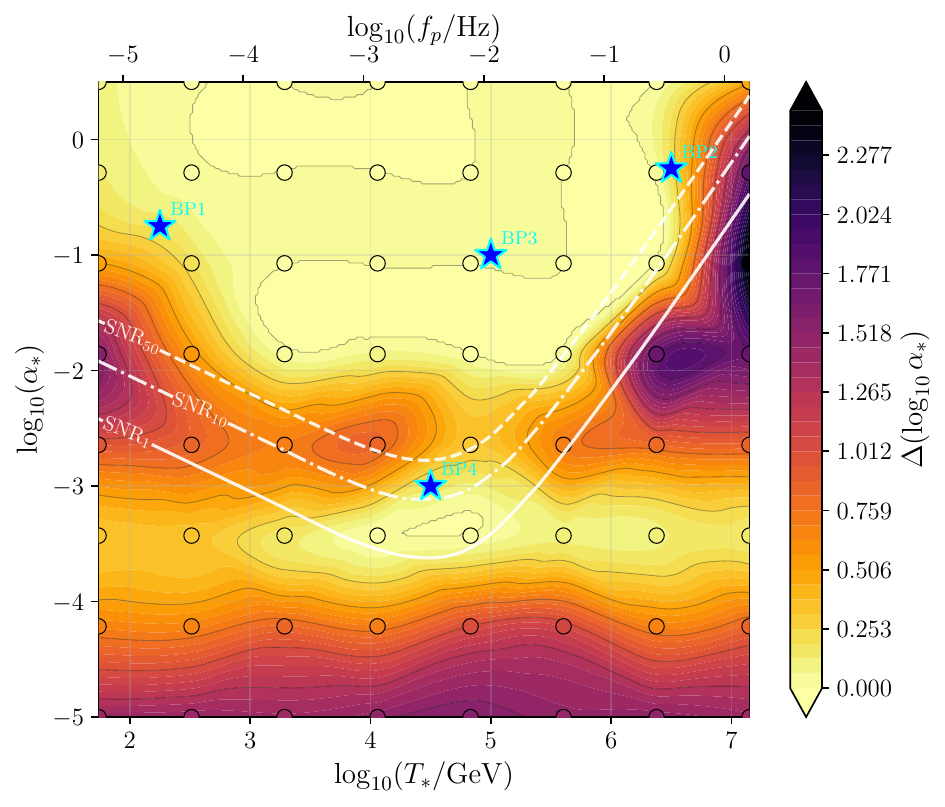}
    \caption{Heat maps of $\Delta(\log_{10}T_*)$ and $\Delta(\log_{10}\alpha_*)$, where $\Delta(\log_{10}X)\equiv |\mathrm{median}(\log_{10}X)-\log_{10}X_{\rm inj}|$ quantifies the offset of the reconstructed median from the injected value. Values $\lesssim 0.1$ correspond approximately to biases below $25\%$. These maps, therefore, complement the posterior-width maps by tracking reconstruction accuracy rather than precision alone.}
    \label{fig:fig7}
\end{figure}

First, while Fig.~\ref{fig:fig1} shows the marginalized uncertainty maps for $\sigma_{T_*}$ and $\sigma_{\alpha_*}$, it is also useful to consider an accuracy diagnostic that explicitly tracks the offset of the reconstructed central value from the injected one. A convenient choice is
\begin{equation}
    \Delta_X \equiv \left| \mathrm{median}(\log_{10}X)-\log_{10}X_{\rm inj}\right|,
\end{equation}
which quantifies the bias of the reconstructed median in logarithmic space. The corresponding maps are shown in Fig.~\ref{fig:fig7}. Ideally, one would like the uncertainty maps in Fig.~\ref{fig:fig1} and the bias maps in Fig.~\ref{fig:fig7} to be simultaneously small. In that case the reconstruction is both precise and accurate.

Second, the heat maps presented here are obtained from 64 injections on an equidistant grid and subsequent interpolation. Although this procedure is already numerically expensive, we expect the resulting maps to be sufficiently accurate for the present forecasting purpose. A denser injection grid would nevertheless be desirable in future work, both to reduce interpolation uncertainties and to obtain still smoother and more precise reconstruction maps.

Third, the instrumental-noise model adopted in this work represents an idealised description of LISA. Although the acceleration-noise amplitude $N_{\rm acc}$ and optical-path noise amplitude $\delta x$ are treated as free nuisance parameters and marginalised over in the Bayesian inference, their frequency dependences are assumed to follow fixed functional forms, with common noise levels assigned to the corresponding instrumental links. In a more realistic configuration, the individual links and test masses may exhibit different noise levels and additional uncertainties in their spectral shapes. Such asymmetries can introduce non-vanishing correlations between the nominally noise-orthogonal $A,E,T$ combinations. More general descriptions, allowing independent noise amplitudes and flexible frequency-dependent spectral shapes, have been investigated in Refs.~\cite{Hartwig:2023pft,Baghi:2023qnq,Muratore:2023gxh}. Incorporating these additional degrees of freedom could modify the correlations between instrumental noise and the domain-wall signal, thereby affecting the achievable reconstruction precision. Our results should therefore be interpreted as baseline forecasts within the adopted symmetric, two-parameter instrumental-noise model.

Closely related to this approximation is our treatment of the LISA $T$ channel. In the idealised equal-arm configuration, the $A,E,T$ combinations diagonalise the instrumental-noise covariance under symmetric noise assumptions, while the gravitational-wave response of the $T$ channel is strongly suppressed at low frequencies. The latter property makes $T$ a useful approximate instrumental-noise monitor, but it does not imply an exactly vanishing gravitational-wave response at arbitrary frequencies. In a realistic constellation, unequal and time-dependent arm lengths modify both the gravitational-wave response and the instrumental-noise transfer functions, potentially introducing residual inter-channel correlations and changing the sensitivity of the $T$ channel \cite{Wang:2020fwa,Karnesis:2021tsh,Du:2025fes}. A treatment based on realistic orbital dynamics and the corresponding time-dependent TDI covariance would therefore constitute an important refinement of the present analysis.

Fourth, the treatment of astrophysical foregrounds represents another relevant modelling assumption. In our baseline analysis, the unresolved DWD foreground is described by a phenomenological broken-power-law spectrum, whose four parameters $(A_1,\alpha_1,A_2,\alpha_2)$ are marginalised over together with the cosmological and instrumental-noise parameters. This prescription allows substantial freedom in the foreground amplitude and spectral shape. However, it does not necessarily encompass the full diversity of spectra predicted by different Galactic binary-population models, including variations in the residual confusion foreground following the subtraction of individually resolvable binaries \cite{Karnesis:2021tsh}. To assess the dependence of reconstruction on the foreground parametrisation, we have additionally considered a more restrictive and commonly used hyperbolic-tangent (tanh) template \cite{Karnesis:2021tsh,Flauger:2020qyi,Caprini:2024hue}, in which the spectral shape is fixed and only the overall amplitude is varied. As discussed in the Appendix \ref{app}, this alternative prescription yields improved reconstruction precision within the assumed model, illustrating the impact of foreground flexibility on the inferred domain-wall parameters. Such an improvement is conditional on the adopted spectral shape adequately describing the actual foreground.

Beyond uncertainties in the frequency dependence, it would also be valuable to go beyond isotropic foreground templates \cite{Giampieri:1997ie,Criswell:2024hfn,Digman:2022jmp,Hindmarsh:2024ttn,Buscicchio:2024wwm}. In the present work, the Galactic DWD foreground is modelled through an effective isotropic, time-averaged spectrum, providing a natural first step for forecasting studies. In reality, the Galactic foreground is anisotropic, and its observed amplitude is modulated by the orbital motion of the LISA constellation. Incorporating this anisotropy and time dependence could provide additional information for separating the Galactic foreground from an approximately isotropic cosmological component. A more comprehensive framework would employ time-dependent detector responses together with population-informed angular templates or a spherical-harmonic description of the foreground. Such an extension could help reduce foreground--signal degeneracies, although its quantitative impact on domain-wall reconstruction would need to be assessed explicitly.

Finally, the reconstruction maps presented here are
conditional on the assumption that the cosmological
SGWB component is generated by domain walls. While
our Bayesian analysis accounts for instrumental noise
and astrophysical foregrounds, it does not attempt
to establish the physical origin of an otherwise
uncharacterised stochastic excess. Other cosmological
mechanisms, including first-order phase transitions
and cosmic strings, may produce spectra that are
partially degenerate with the domain-wall signal
over the accessible LISA frequency range. Distinguishing between these possibilities would
require a comparison of competing cosmological
hypotheses through Bayesian evidence calculations \cite{Datta:2026udh}
or other model-selection techniques, potentially
including transdimensional approaches such as
reversible-jump Markov chain Monte Carlo
\cite{Puecher:2026poi}. A further extension would involve joint inference
allowing multiple cosmological SGWB components
to coexist. Such analyses could introduce additional
parameter degeneracies and modify the inferred
domain-wall parameter uncertainties. The present results should therefore be understood
as conditional parameter-reconstruction forecasts,
rather than as a demonstration of unique domain-wall
source identification. Extending the analysis towards
realistic instrumental responses, anisotropic and
time-dependent foregrounds, and cosmological model
discrimination represents a natural direction for
future work.

\section{Conclusion}
\label{sec:conclusion}

In this work, we have investigated the Bayesian
reconstruction of peaked domain-wall gravitational-wave
signals in the LISA band and constructed reconstruction
maps over the signal-parameter plane. Our objective
was to identify the regions in which the underlying
domain-wall parameters can be inferred with both
small posterior uncertainties and weak parameter
degeneracies, going beyond the conventional assessment
of signal detectability. The resulting maps provide
a quantitative characterization of the reconstruction
capabilities of LISA under the adopted instrumental-noise
and astrophysical-foreground modelling assumptions.

Our analysis demonstrates that the reconstruction
capability is closely related to the position of
the spectral turnover relative to the LISA sensitivity
window. When the peak lies well below the detector
band, the observed signal is dominated by its
ultraviolet tail, and different combinations of
domain-wall parameters can produce nearly
indistinguishable spectra. A similar degeneracy
arises when the peak lies well above the detector
band and only the infrared tail is accessible.
In contrast, when the spectral turnover lies
within or near the sensitive frequency range,
the observed spectrum contains information
about both the overall amplitude and the
spectral curvature, allowing the underlying
parameters to be constrained more independently.

To characterize these features, we have employed
principal-component diagnostics of the posterior
covariance, focusing on the parameter 
$\kappa=\lambda_\parallel/\lambda_\perp$
and the smaller eigenvalue $\lambda_\perp$.
These quantities provide complementary information
about posterior anisotropy and the constraining
power along different parameter directions.
Importantly, a $\kappa$ does not
necessarily imply successful reconstruction,
since weak signals may yield nearly isotropic
but broad, prior-dominated posteriors.
A reliable assessment therefore requires
the simultaneous examination of posterior
geometry, marginalised uncertainties, and
signal-to-noise ratio.

For the baseline foreground prescription,
we identify a favourable reconstruction
region corresponding approximately to
annihilation temperatures
$10^3\text{--}10^6\,\mathrm{GeV}$.
Within this region, sufficiently strong
signals, typically with ${\rm SNR}\gtrsim 50$,
allow both domain-wall parameters to be
reconstructed with comparatively small
uncertainties and reduced degeneracies.
We also find that weaker signals with
${\rm SNR}\sim 10$ can be effectively
reconstructed, although only within
a more restricted region concentrated
near $T_*\sim 10^{5-6}\,\mathrm{GeV}$.
In terms of observable spectral properties,
this weaker-signal regime is associated
approximately with peak amplitudes satisfying
$4\times10^{-12}\lesssim
\Omega_{\rm GW}^{\rm peak}h^2
\lesssim 4\times10^{-11}$
and peak frequencies in the range
$30\,{\rm mHz}\lesssim f_p
\lesssim 100\,{\rm mHz}$.

An important feature of the favourable
reconstruction region is that the marginalised
uncertainties in the original domain-wall
parameters become comparable to the
corresponding principal-component widths.
This indicates that the improved reconstruction
is not merely associated with a reduction
in posterior elongation, but reflects
the ability to constrain both physical
parameters simultaneously.
The peak-in-band configuration therefore
differs qualitatively from tail-dominated
reconstruction scenarios \cite{Datta:2026fav},
where the observable spectrum constrains
primarily a single effective combination
of the underlying parameters.

We have also investigated the sensitivity of the
reconstruction results to the adopted Galactic
foreground prescription. Alongside the flexible
four-parameter broken-power-law model used in
the baseline analysis, we considered a more
restrictive hyperbolic-tangent template
(see Appendix~\ref{app}), in which the spectral
shape is fixed and only the overall amplitude
is varied. This alternative prescription
generally improves reconstruction precision,
extending the region in which weaker domain-wall
signals can be effectively constrained.
The improvement reflects the reduced freedom
of the Galactic foreground to mimic variations
in the cosmological signal spectrum.

Interestingly, the improved precision does not
necessarily translate into reduced posterior
anisotropy. The corresponding condition-number
maps indicate a somewhat narrower region of
weak parameter degeneracy under the tanh
prescription, illustrating that tighter
posterior constraints do not always imply
better parameter identifiability.
This comparison highlights the complementary
roles of posterior uncertainties and
principal-component diagnostics in assessing
reconstruction quality. We emphasise that
the improvement obtained with the more
restrictive foreground template remains
conditional on the assumed spectral shape
adequately describing the actual Galactic
foreground.

Finally, our results should be interpreted
within the scope of the modelling assumptions
adopted in this work.
The analysis employs an idealised equal-arm
LISA configuration with symmetric
instrumental-noise properties and fixed
noise spectral shapes, while marginalising
over the corresponding noise amplitudes.
The astrophysical foregrounds are described
through effective isotropic, time-independent
spectra, without explicitly accounting
for the annual modulation associated with
the anisotropic Galactic binary population.
Moreover, the reconstruction forecasts
are conditional on the assumption that
the cosmological stochastic background
follows only the domain-wall spectral template.
Extending the present framework to include
more realistic instrumental responses,
time-dependent foreground anisotropies,
and discrimination among competing
cosmological sources constitutes an
important direction for future studies.

Overall, our analysis identifies the
regions of the domain-wall parameter
space in which LISA can achieve genuine
two-parameter reconstruction, rather
than merely detect a stochastic excess.
The resulting reconstruction maps provide
a useful quantitative benchmark for
assessing the observational prospects
of particle-physics scenarios that
generate peaked gravitational-wave
spectra in the milli-Hz frequency band.

\section*{Acknowledgments}
The work of SD is supported by the Slovenian Quantum Science Hub co-funded by the Marie Sk{\l}odowska-Curie Actions programme (GA-101177446) and the Slovenian Research and Innovation Agency (ARIS), contract number 5110-18/2025-5. 
The work of RS is supported by the research project TAsP (Theoretical Astroparticle Physics) funded by the Istituto Nazionale di Fisica Nucleare (INFN). We thank Sabir Ramazanov and Alexander Vikman for pointing out useful references. 
\appendix

\section{Reconstruction dependence on Galactic foreground templates}
\label{app}

In this appendix, we investigate the sensitivity of the
domain-wall parameter reconstruction to the assumed
spectral form of the unresolved Galactic double-white-dwarf
(DWD) foreground. The baseline analysis employs a flexible
phenomenological broken-power-law parametrisation with four
free parameters, $(A_1,\alpha_1,A_2,\alpha_2)$, which are
jointly inferred with the cosmological and instrumental-noise
parameters. To assess the impact of this modelling freedom,
we consider an alternative prescription in which the
frequency dependence of the Galactic foreground is fixed
and only its overall amplitude is allowed to vary.

Specifically, we adopt the hyperbolic-tangent (tanh)
foreground template
\cite{Caprini:2024hue,Karnesis:2021tsh,Samanta:2025jec},
given by
\begin{equation}
    h^2 \Omega_{\rm Gal}^{\rm GW}(f)
    =
    \frac{f^3}{2}
    \left(\frac{f}{1\,{\rm Hz}}\right)^{-7/3}
    \left[
    1+
    \tanh\left(
    \frac{f_{\rm knee}-f}{f_2}
    \right)
    \right]
    \exp\!\left[
    -\left(\frac{f}{f_1}\right)^\nu
    \right]
    h^2\Omega_{\rm Gal}.
    \label{gal_tanh}
\end{equation}
Here, $h^2\Omega_{\rm Gal}$ denotes the foreground
normalisation and is treated as the only free Galactic
foreground parameter in the Bayesian inference (see Table~\ref{tab:tanh_parameters} for details).
The characteristic frequencies $f_1$ and $f_{\rm knee}$
are determined by the observation time through
\begin{align}
    \log_{10}\left(\frac{f_1}{{\rm Hz}}\right)
    &=
    a_1
    \log_{10}\left(
    \frac{T_{\rm obs}}{{\rm year}}
    \right)
    +b_1,
    \\
    \log_{10}\left(\frac{f_{\rm knee}}{{\rm Hz}}\right)
    &=
    a_k
    \log_{10}\left(
    \frac{T_{\rm obs}}{{\rm year}}
    \right)
    +b_k,
\end{align}
where the coefficients are fixed to
\begin{align}
    a_1 &= -0.15,
    &
    b_1 &= -2.72,
    \nonumber\\
    a_k &= -0.37,
    &
    b_k &= -2.49,
    \nonumber\\
    \nu &= 1.56,
    &
    f_2 &= 6.7\times 10^{-4}\,{\rm Hz}.
\end{align}

\begin{table}[htbp]
\centering
\begin{tblr}
	{
    hlines,
    vlines,
    row{1} = {bg=gray7, fg=white, font=\bfseries},
column{1} = {bg=gray9},
cell{1}{1} = {bg=gray7, fg=white},
}

\textbf{Parameter} & \textbf{Injected value} & \textbf{Prior (uniform)} \\
$\log_{10}(h^2\Omega_{\rm Gal})$      & $-12.0$ & $(-14.0,-10.0)$ \\
\end{tblr}
\caption{{$\tanh$} foreground normalisation fiducial value, and uniform prior range used in the inference.}
\label{tab:tanh_parameters}
\end{table}

Unlike the broken-power-law prescription adopted in the
main analysis, the tanh template does not introduce
independent nuisance parameters controlling the foreground
slopes and turnover. It therefore provides a useful
comparison for assessing how the assumed knowledge of
the Galactic spectral shape influences the reconstruction
of the domain-wall parameters.
\begin{figure}
    \centering
   \includegraphics[width=0.5\linewidth]{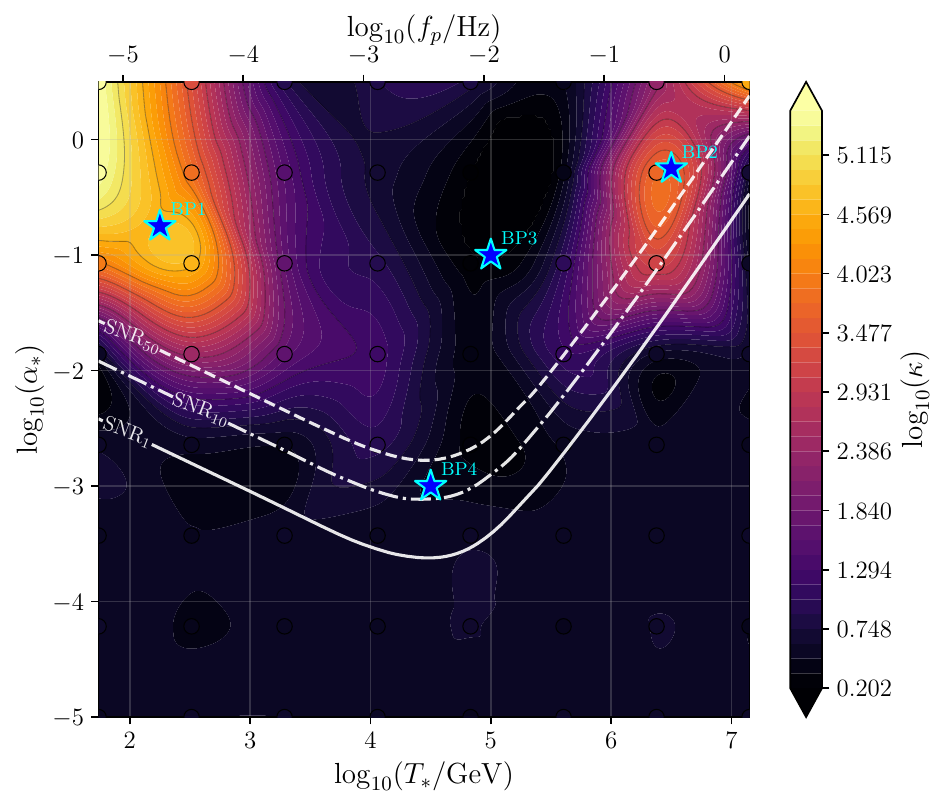}\includegraphics[width=0.5\linewidth]{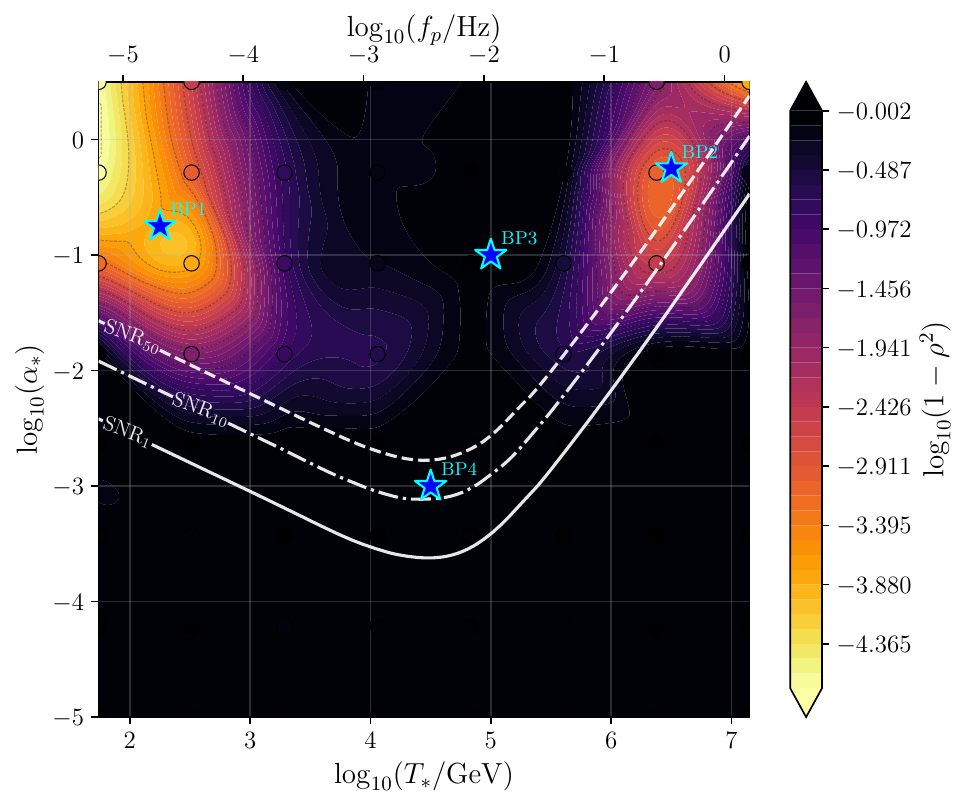}\\
    \includegraphics[width=0.5\linewidth]{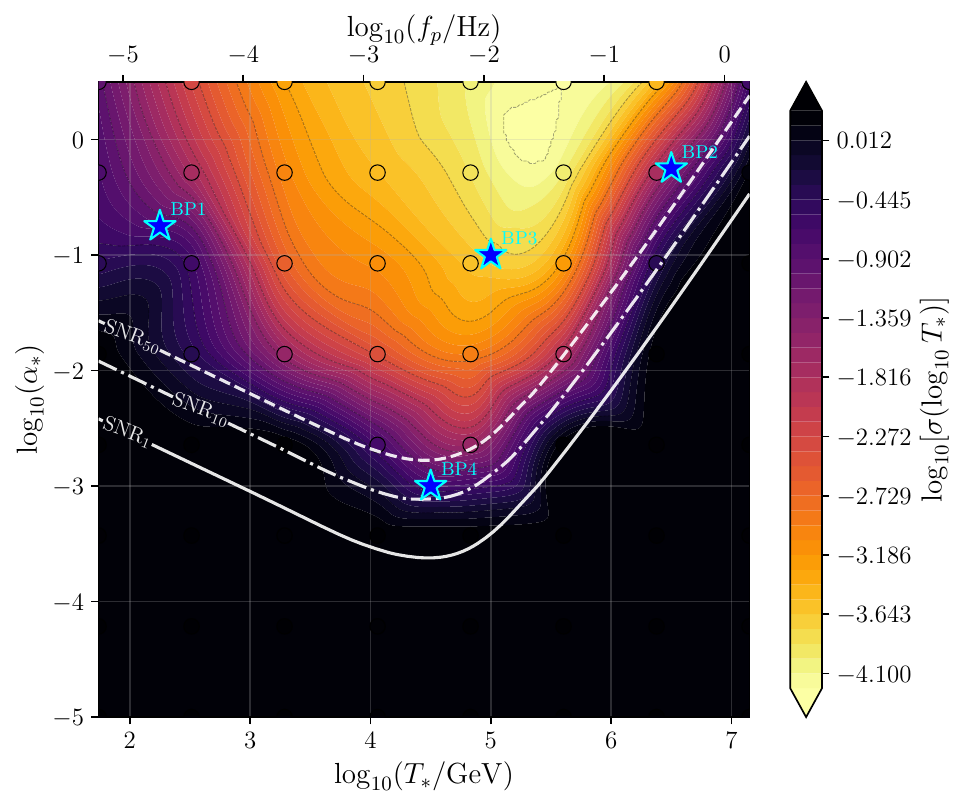}\includegraphics[width=0.5\linewidth]{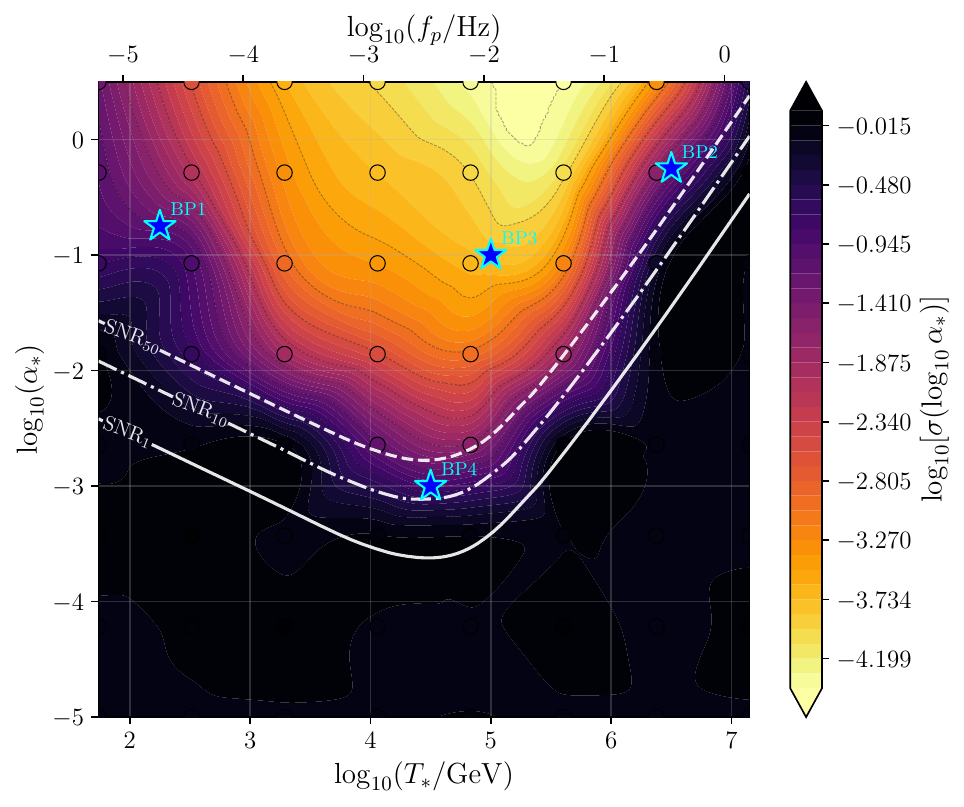}
    \caption{Description same as Fig. \ref{fig:fig1}, but with the foreground model defined in Eq.\eqref{gal_tanh}.}
    \label{fig:figap1}
\end{figure}
\begin{figure}
    \centering
    \includegraphics[width=0.5\linewidth]{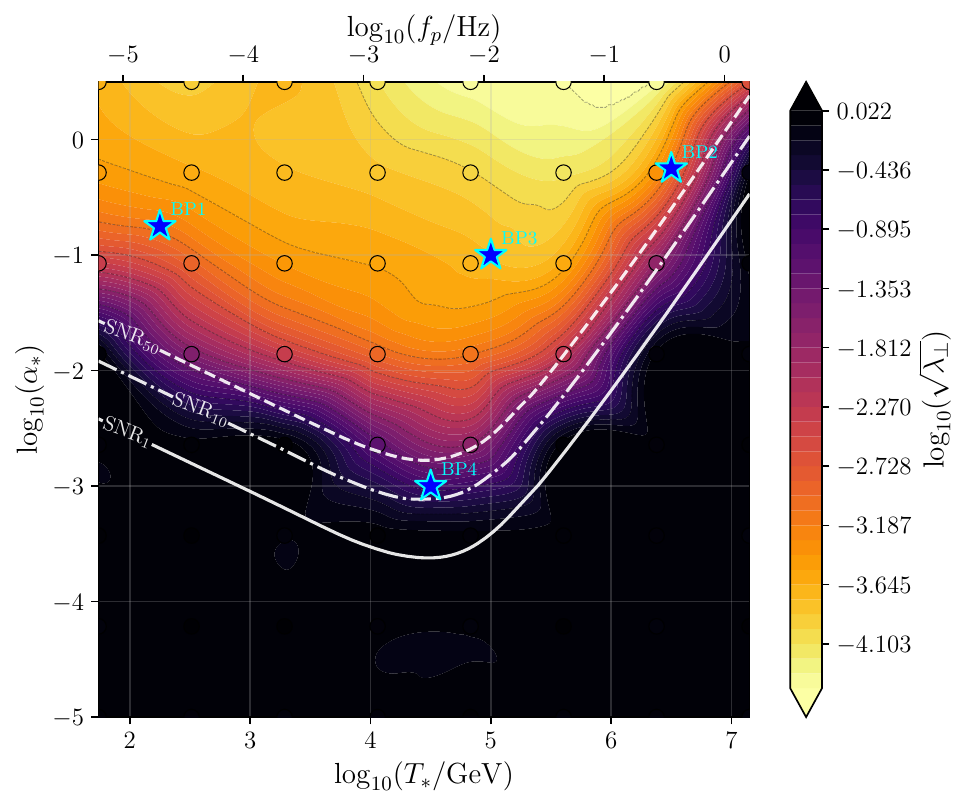}\includegraphics[width=0.5\linewidth]{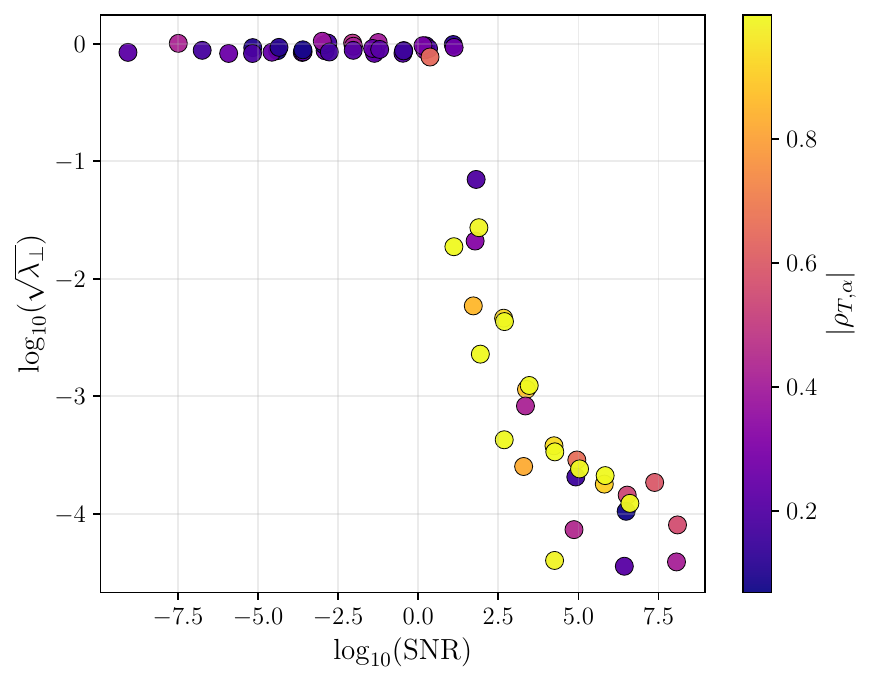}
    \caption{Description same as Fig. \ref{fig:fig2}, but with the foreground model defined in Eq.\eqref{gal_tanh}.}
    \label{fig:figap2}
\end{figure}
To assess the impact of the foreground prescription,
we reproduce the reconstruction maps of the main
analysis in Figs.~\ref{fig:figap1} and
\ref{fig:figap2}, now adopting the tanh template
of Eq.~\eqref{gal_tanh}. The injection grid,
domain-wall signal model, instrumental-noise
parametrisation, and Bayesian inference procedure
are kept unchanged, allowing a direct comparison
between the two foreground choices.

The resulting maps generally exhibit improved
reconstruction precision relative to the more
flexible foreground treatment, with regions of
small posterior uncertainties extending towards
weaker domain-wall signals. This improvement can
be attributed to the reduced freedom of the
Galactic foreground to mimic variations in the
cosmological spectrum. In the baseline analysis,
changes in the foreground amplitude, slopes,
and turnover can partially compensate for
variations in the domain-wall parameters,
thereby broadening their marginalised posterior
distributions. Fixing the foreground spectral
shape reduces these degeneracies and leads
to tighter parameter constraints.

Interestingly, the improved precision does
not necessarily translate into a corresponding
reduction in posterior anisotropy. The
$\kappa$ maps indicate a somewhat
narrower region of weak parameter degeneracy
under the tanh prescription. This reflects
the fact that the constraints along the two
principal posterior directions need not
improve by the same relative amount.
Consequently, tighter posterior distributions
can still exhibit larger values of $\kappa$.

We emphasise that the improvement obtained
with the tanh template is conditional on its
assumed spectral shape adequately describing
the true Galactic foreground. The more
restrictive parametrisation does not
necessarily constitute a more realistic
foreground model, but instead illustrates
the reconstruction capabilities achievable
under stronger assumptions about the
foreground spectrum. This comparison
therefore highlights the sensitivity of
domain-wall reconstruction forecasts to
foreground modelling choices and reinforces
the importance of considering both posterior
uncertainties and parameter-degeneracy
diagnostics.
\bibliography{bibliography}
\end{document}